\documentclass[12pt]{iopart}
\usepackage{iopams}  

\newcommand{\Katrin}{\textsc{Katrin}}

\newcommand{\KEMField}{\textsc{KEMField}}
\newcommand{\Kassiopeia}{\textsc{Kassiopeia}}

\usepackage{graphicx}
\usepackage{subfigure}
\usepackage{epstopdf}
\usepackage{booktabs}
\usepackage{float}
\usepackage{hyperref}

\usepackage{color}
\usepackage{listings}
\definecolor{Maroon}{rgb}{0.5,0,0}
\definecolor{DarkOliveGreen}{rgb}{0,0.5,0}
\lstdefinelanguage{XML}
{
  basicstyle=\ttfamily\footnotesize,
  morestring=[b]",
  moredelim=[s][\bfseries\color{Maroon}]{<}{\ },
  moredelim=[s][\bfseries\color{Maroon}]{</}{>},
  moredelim=[l][\bfseries\color{Maroon}]{/>},
  moredelim=[l][\bfseries\color{Maroon}]{>},
  morecomment=[s]{<?}{?>},
  morecomment=[s]{<!--}{-->},
  commentstyle=\color{DarkOliveGreen},
  stringstyle=\color{blue},
  identifierstyle=\color{red},
	showstringspaces=false,
	tabsize=4
}

\begin{document}

\title{Kassiopeia{}: A Modern, Extensible C++ Particle Tracking Package}

\author{Daniel Furse$^1$, Stefan Groh$^2$, Nikolaus Trost$^3$, Martin Babutzka$^2$, 
John P. Barrett$^1$, Jan Behrens$^4$, Nicholas Buzinksy$^1$, Thomas Corona$^{5,6}$, Sanshiro Enomoto$^7$, Moritz Erhard$^2$, Joseph A. Formaggio$^1$, Ferenc Gl\"uck$^{3,8}$, Fabian Harms$^3$, Florian Heizmann$^2$, Daniel Hilk$^2$, Wolfgang K\"afer$^2$, Marco Kleesiek$^2$, Benjamin Leiber$^2$, Susanne Mertens$^9$, Noah S. Oblath$^1$, Pascal Renschler$^2$, Johannes Schwarz$^3$, Penny L. Slocum$^{10}$, Nancy Wandkowsky$^3$, Kevin Wierman$^{5,6}$ and Michael Zacher$^4$}

\address{$^1$ Laboratory for Nuclear Science, Massachusetts Institute of Technology, Cambridge, MA 02139, USA}
\address{$^2$ Institute of Experimental Nuclear Physics (IEKP), Karlsruhe Institute of Technology (KIT), Wolfgang-Gaede-Str. 1, 76131 Karlsruhe, Germany}
\address{$^3$ Institute for Nuclear Physics (IKP), Karlsruhe Institute of Technology (KIT), Hermann-von-Helmholtz-Platz 1, 76344 Eggenstein-Leopoldshafen, Germany}
\address{$^4$ Institut f\"ur Kernphysik, Westf\"alische Wilhelms-Universit\"at M\"unster, 48149 M\"unster, Germany}
\address{$^5$ Department of Physics and Astronomy, University of North Carolina, Chapel Hill, NC 27599, USA}
\address{$^6$ Triangle Universities Nuclear Laboratory, Durham, NC 27708, USA}
\address{$^7$ Center for Experimental Nuclear Physics and Astrophysics, Department of Physics, University of Washington, Seattle, WA 98195, USA}
\address{$^8$ Wigner Research Institute for Physics, Budapest POB 49, Hungary}
\address{$^9$ Max Planck Institute for Physics, Munich \& Technical University Munich, 80333 Munich, Germany}
\address{$^{10}$ Department of Physics, Yale University, PO Box 208120, New Haven, CT 06520, USA}

\ead{nikolaus.trost@kit.edu}
\vspace{10pt}
\begin{indented}
\item[]December 2016
\end{indented}

\begin{abstract}
The \Kassiopeia{} particle tracking framework is an object-oriented software package using modern C++ techniques, written originally to meet the needs of the \Katrin{} collaboration. \Kassiopeia{} features a new algorithmic paradigm for particle tracking simulations which targets experiments containing complex geometries and electromagnetic fields, with high priority put on calculation efficiency, customizability, extensibility, and ease of use for novice programmers.
To solve \Kassiopeia{}'s target physics problem the software is capable of simulating particle trajectories governed by arbitrarily complex differential equations of motion, continuous physics processes that may in part be modeled as terms perturbing that equation of motion, stochastic processes that occur in flight such as bulk scattering and decay, and stochastic surface processes occuring at interfaces, including transmission and reflection effects. This entire set of computations takes place against the backdrop of a rich geometry package which serves a variety of roles, including initialization of electromagnetic field simulations and the support of state-dependent algorithm-swapping and behavioral changes as a particle's state evolves. Thanks to the very general approach taken by \Kassiopeia{} it can be used by other experiments facing similar challenges when calculating particle trajectories in electromagnetic fields. It is publicly available at \url{https://github.com/KATRIN-Experiment/Kassiopeia}.
\end{abstract}

%\submitto{\NJP}
%\maketitle

\section{Introduction}

\Kassiopeia{} is a software package for the purpose of tracking particles in complex geometries and fields. It has been developed in order
to meet the simulation needs of the KATRIN collaboriation which endeavours to measure the absolute neutrino mass scale through tritium $\beta$-decay.
Strong evidence for the existence of non-zero neutrino mass follows from the legion of experiments demonstrating flavor oscillation phenomena \cite{Fukuda1998, Ahmad2001, Ahmad2002, Aharmim2008, Eguchi2003, An2012, Abe2011}. The discovery of neutrino oscillations (hence, neutrino mass) is the first demonstration of neutrino properties beyond the Standard Model prescription. However, oscillation phenomena depend only on the differences of the squares of neutrino mass eigenvalues $\Delta m_{ij}^2 \equiv m_j^2 - m_i^2$; the absolute neutrino mass scale does not enter into the description of oscillation phenomena. As such, the absolute neutrino mass remains one of the foremost open questions in neutrino physics at the present time.

The most sensitive direct searches for the electron neutrino mass
to date are based on the investigation of the electron spectrum
of tritium $\beta$-decay. The electron energy spectrum of tritium $\beta$-decay for a neutrino with
component masses $m_1, m_2,$ and $m_3$ (with mixing amplitudes $U_{e1}, U_{e2}$, and $U_{e3}$, respectively) is given (with some simplifications\footnote{Note that we have neglected modifications of the energy spectrum due to the nuclear matrix element and the molecular final state distribution.}) by:
\begin{equation}
\mkern-36mu \frac{dN}{dE} \propto F(Z,E) p E(E_0-E) \sum_{i=1,3}|U_{ei}|^2[(E_0-E)^2-m_i^2]^{\frac{1}{2}}\Theta (E_0-E-m_i), \label{mother}
\end{equation}
where $E$ (p) denotes the electron's kinetic energy (momentum), $E_0$ corresponds to the total decay energy, $F(Z,E)$
is the Fermi function, taking into account the Coulomb interaction
of the outgoing electron in the final state, and $\Theta (E_0-E-m_i)$ is the step function that ensures energy conservation. As both the matrix elements and $F(Z,E)$ are independent
of $m_\nu$, the dependence of the spectral shape on $m_\nu$ is
given by the phase space factor only. The bound on
the neutrino mass from tritium $\beta$-decay is independent of whether
the electron neutrino is a Majorana or a Dirac particle.

Although the history of beta spectroscopy spans a variety of different magnetic and electrostatic spectrometers, the technique which has demonstrated the greatest sensitivity to neutrino mass has been MAC-E-Filters (Magnetic Adiabatic Collimation with Electrostatic Filtering).  This type of spectrometer, originally based on the work by Kruit \cite{Kruit1983} was later utilized by the Mainz~\cite{Kraus2005} and Troitsk~\cite{Aseev2011} experiments to
set a limit on the neutrino mass on the level of 2 eV. The MAC-E filter technique demands a smoothly varying magnetic field
and has an energy resolution which is dictated by the ratio of the minimum to maximum magnetic field strength. This dictates that 
better energy resolution be accompanied by an increase in the physical size of the spectrometer. 
The KATRIN experiment's massive size and other design optimizations \cite{kdr:2004} will allow it to reach an energy 
resolution of 0.93 eV, and should allow it to place a limit on the neutrino mass approximately
an order of magnitude better than the current state of the art \cite{Kraus2005,Aseev2011}.

\begin{figure*}[t]
\begin{center}
\includegraphics[width=17cm]{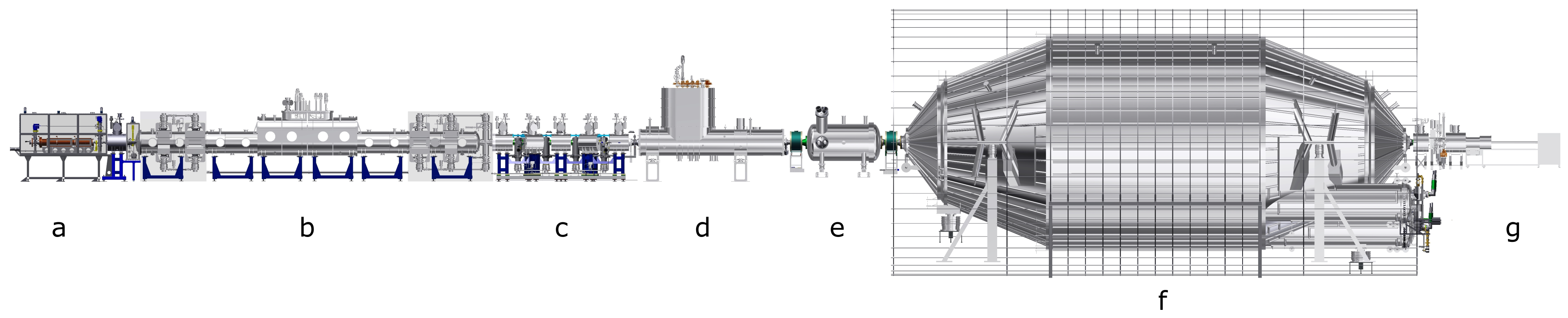}
\caption{Schematic overview of the 70-m KATRIN experimental beam-line: (a) the rear section, (b) the tritium source, (c) the differential-pumping section, (d) the cryogenic-pumping section, (e) the pre-spectrometer, (f) the main spectrometer and air-coil framework, and (g) the focal-plane detector system.}
\label{fig:KATRIN}
\end{center}
\end{figure*}

 Figure~\ref{fig:KATRIN} illustrates the overall components of the experiment, which include: (a) a rear section, used for calibrating the response of the detector and monitoring the source strength, (b) a Windowless Gaseous Tritium Source (WGTS), where $10^{11}$ electrons are produced per second by the decay of molecular tritium gas at a temperature of 30~K, (c) an electron transport and tritium elimination section, comprising of (c) an active differential pumping (DPS) followed by (d) a passive cryo-pumping section (CPS), where the tritium flow is reduced by more than 14 orders of magnitude, (e) the electrostatic pre-spectrometer which offers the option to pre-filter the low-energy part of the tritium decay spectrum, (f) the large electrostatic main spectrometer of MAC-E-filter type which represents the precision energy filter for electrons, and (g) a segmented Si-PIN diode array to count the transmitted electrons. Further details on the experiment's design and performance parameters can be found elsewhere~\cite{kdr:2004}. 

A suitable computational model of such a complex experiment is indeed necessary if one is to properly assess the results obtained by the experiment.  
Such a tool is essential for many tasks, from the estimation of background rates and systematic effects to modeling signal electron energy loss and backscattering at the silicon detector. Ultimately, detailed simulations based on measurements performed during installation are the most accurate tool for evaluating the sensitivity of the entire experiment, giving them a position of central importance.  However, such performance goals often impose strict (and often contradictory) conditions on the simulation software.  For one, the software needs to be able to accurately propagate electrons through the complex fields and geometries found in the experiment. while at the same time retain high efficiency and flexibility for such propagation.  In addition, KATRIN needs to be able not just to calculate such trajectories, but also compute the electromagnetic fields {\em in situ}.  Though tools currently exist for each task separately ({\it e.g.} GEANT4~\cite{GEANT4} for particle tracking, COMSOL~\cite{COMSOL} for electromagnetic field calculations), none appeared adequate for the joint task.  Tracking and navigation algorithms designed without this situation in mind often suffer from unacceptably long computation times, which are typically compounded further in navigationally complex situations where stepping lengths are on the order of the dimensions of geometrical components. 

Furthermore, neither software provides the required flexibility sought by the KATRIN experiment. For instance, a researcher trying to design a new component of the spectrometer electrode system needs quick feedback on its influence on the fields, and ultimately needs to understand the influence it has on the dynamical properties of the spectrometer. Another might need to understand tritium ion propagation and scattering in the source of the experiment. Even two people working on the same topic might be interested in completely different aspects of the same physics, requiring different output from an otherwise identical simulation. Such examples are inexhaustible, which indicates a need for a simulation package that is very granular and allows for a large space of possible module combinations and arrangements in a user configurable way.  These facts combined with our requirements for modularity, extensibility, and ease-of-use for 
novice programmers unfortunately have ruled out the majority of existing candidates.

In light of these facts, development began on \textsc{Kassiopeia}, a new general particle and field simulation package designed to meet the diverse needs outlined above. Since modularity, encapsulation and speed are essential, the development team decided to implement the simulation as an object-oriented design written in C++, with a user interface based primarily on configuration files written in XML. We believe that the current version of the code, which is now 
publicly available, broadly satisfies the requirements set out above and will present a valuable and complementary approach to particle simulation packages available to the wider physics community.

The organization of this paper is as follows. Section \ref{sec:design} details the overlying design of the \textsc{Kassiopeia} software package and its configuration. Sections \ref{sec:geometry} and \ref{sec:fields} describe the geometry model and the electromagnetic field calculations respectively. Sections \ref{sec:generation} through \ref{sec::termination} provide a description of the generation of particle states, their propagation through simulation 
geometries by solving of their equations of motion, the treatment of stochastic interactions, and termination. Of subsequent interest are: section \ref{sec:output} which outlines the various data output options, section \ref{sec:navigation} which describes the command structure responsible for maintaining the simulation state machine, and section \ref{sec:visualization} which demonstrates the visualization capabilities of this software. Finally, we conclude with some example use-cases and validation in sections \ref{sec:validation} and \ref{sec:conclusion}.

\section{General design}
\label{sec:design}

The goal of each particle tracking software and therefore also of \textsc{Kassiopeia} is to simulate the evolution of the physical state of multiple particles with very high precision and efficiency. 
The particle therefore represents a fundamental object, whose properties are to be modified by the algorithms of the simulation software. The inherent properties of a particle, its mass $m$ and electric charge $q$, are fixed during initialization, while the dynamic properties, such as its position $\mathbf{x}$, and momentum $\mathbf{p}$, will evolve as the simulation progresses.

%The evolution of multiple particles has to be organized into a specific data structure, which needs to be filled by the simulation, as explained in the next paragraph.
%Afterwards the basic simulation workflow will be detailed so that at the end of this section the modularity of the simulation algorithm can be presented, which is one of the main powerful features of \textsc{Kassiopeia}.

\subsection{Organizational structure}

\begin{figure}[tbp]
 \centering
    \includegraphics[width=0.9\textwidth]{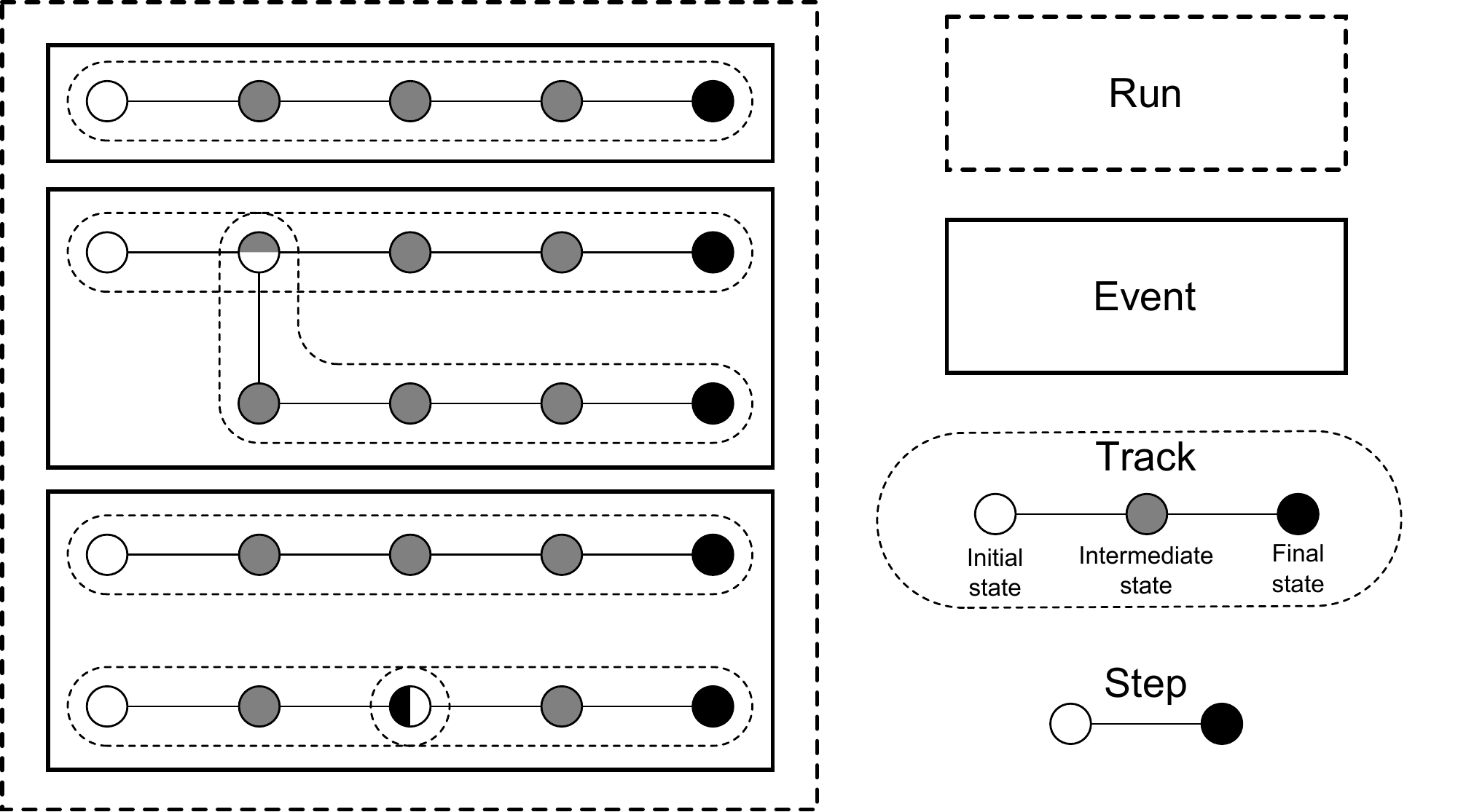}
 \caption[Organization structure of a run]{Schematic representation of a run with 3 events and a total of 6 tracks. Each track starts with an initial state (white) and ends with a final state (black). It consists of multiple steps - four for most of the given examples. Via interaction processes such as ionization within a track, a new particle and therefore a new track can be generated (track 3). Thus an event can consist out of multiple tracks, which is also the case for an event of a radioactive decay chain for example, creating multiple initial particles and therefore multiple tracks within one event (track 4 and 5). A track can be split if it crosses a surface or changes a geometrical space, thus ending the old track and starting a new one (track 5 and 6). Figure from \cite{groh:2015}.}
\label{fig:design:runeventtrackstep}
\end{figure}

\textsc{Kassiopeia}'s data structure, which is filled by output from the simulation, is organized into four intuitive levels of detail: Step, Track, Event and Run. A schematic representation of this classification is visualized in figure \ref{fig:design:runeventtrackstep}. The individual levels are detailed as follows:

\begin{itemize}
	\item \textbf{Step:} The lowest level of organization in the simulation is a step. It represents the evolution of a particle over a small amount of time and space from an initial to a final state. The propagation of the particle is achieved by solving the equations of motion and by considering a variety of interactions with the surrounding matter and fields. Additionally, navigation within the defined geometry is performed to detect the crossing of surfaces or space boundaries.
	\item \textbf{Track:} The complete evolution of a particle from its point of origin to its termination is called a track, which can be seen as a sequential collection of steps. A particle and therefore a track, is typically created within an event generator or through an interaction like ionization. It can be terminated by a collection of terminators depending on specific states of the simulation. Additionally, a particle can also be terminated and a new one generated by the navigation when crossing a surface or changing a space, which thereby splits the track into two.
	\item \textbf{Event:} The next level of organization is an event, which is a collection of causally related tracks. Each event typically has one primary track corresponding to the primary particle created by a generator, and optionally additional secondary tracks created by splitting of the primary track or by new particles being generated during an interaction process. There are also specific generators which produce multiple causally related primary particles, for example in a radioactive decay sequence. Within one event, the primary particles created and all of their descendants are tracked step-by-step until they are terminated.
	\item \textbf{Run:} The highest level of organization within \textsc{Kassiopeia} is the run, which is a collection of events, whose number is pre-defined by the user in the configuration file. It represents one execution of the simulation for a fixed experimental setup. Multiple runs can be realized by running multiple \textsc{Kassiopeia} instances and merging the produced output files at the end.
\end{itemize}

\subsection{Simulation work-flow}

\begin{figure}[tbp]
 \centering
    \includegraphics[width=0.9\textwidth]{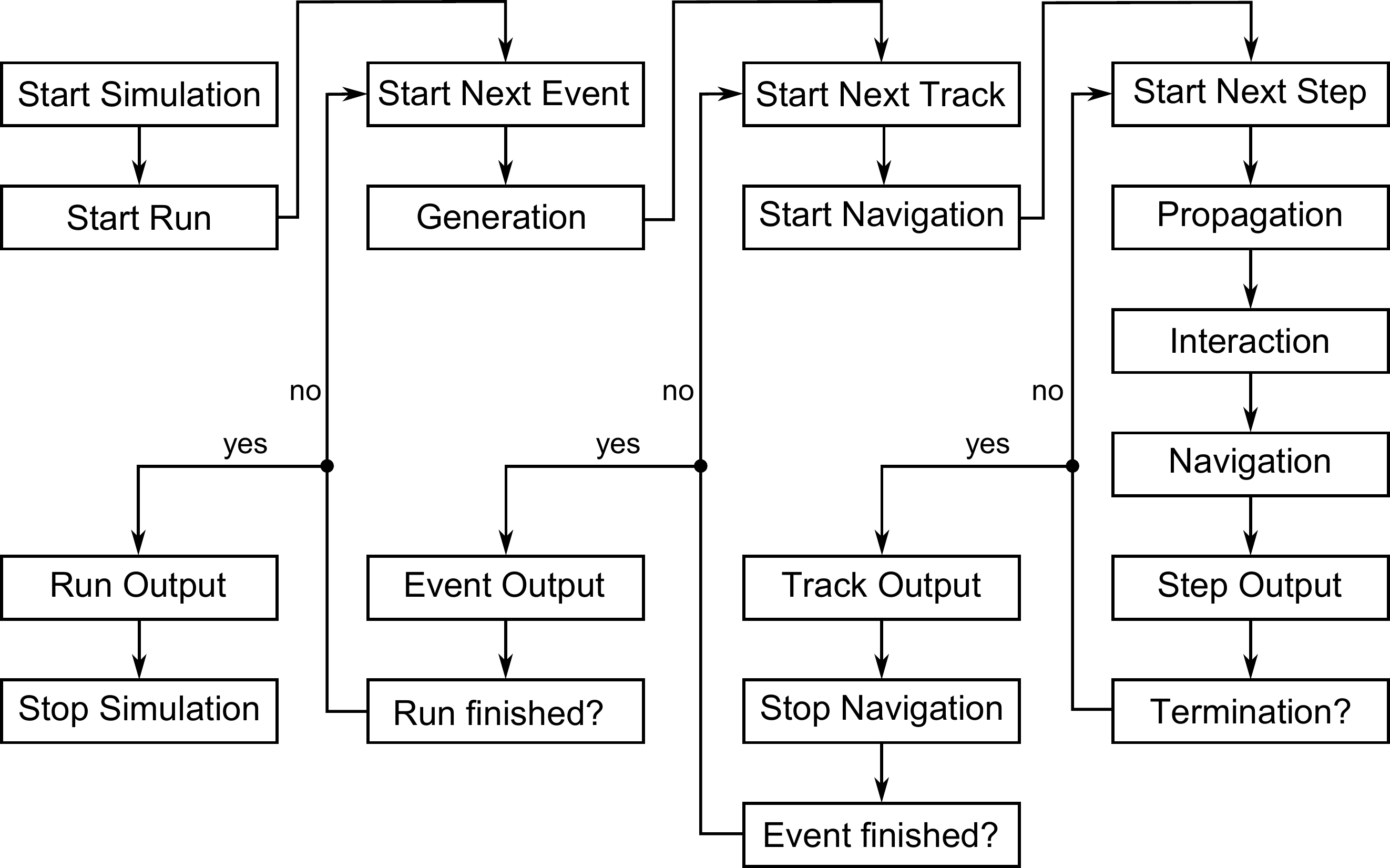}
 \caption[Flow chart of the simulation algorithm]{Simplified schematic representation of \textsc{Kassiopeia}'s simulation algorithm, composed out of three loops over events, tracks and steps. Figure from \cite{groh:2015}.}
\label{fig:design:flowchart}
\end{figure}

The introduced data structures need to be populated by the simulation algorithm. A simplified and schematic chart of the simulation work flow is visualized in figure \ref{fig:design:flowchart}. 
When the simulation is started, first the XML configuration file is parsed and the defined objects used in the simulation will be built and initialized as will be detailed in section \ref{sec:initialization}. Then the event loop is executed $n$ times and in each loop a user-specified generator will produce one or multiple initial particles. For each of these particles a track is created and consecutive steps are performed until the track is terminated. User-defined quantities of the track including the initial and final particle state can then be written to disk before the next track is executed. If the tracking of all particles of the event is completed, including secondary particles created within the tracks, the specific event being executed is finished and the corresponding event output is written. After all $n$ events have been completed, the run output is written and the simulation ends after the deinitialization of all created objects.

The most important part during particle tracking is the step loop, which is typically repeated a large number of times for each track. The schematic representation displayed in figure \ref{fig:design:flowchart} corresponds to a simplification of a more sophisticated algorithm. In each step the particle is propagated by integrating its equations of motion over the user-defined step size. This is typically the most expensive part of the simulation as it involves many calculations of complex electric and magnetic fields, gradients and potentials. After the propogation step has been evaluated, the particle's mean free path length for each of the given interaction processes is calculated and the length at which the process will occur is determined probabilistically. If the randomly
generated interaction length is less than the length of particle step's trajectory, then the terminal position of the particle is adjusted accordingly and the interaction process is executed on the modified final state of the particle. Additionally, the navigator checks if the particle has crossed any geometrical boundaries within the calculated trajectory. If this is the case, the final state of the particle is adjusted to lie on the crossing position of its trajectory with the given geometrical boundary. Interactions or commands associated with the boundary may then modify the particle state or induce a change in the configuration of the simulation, as will be detailed in the next section.  
%If the particle's position is on a surface on the beginning of the step, only the interaction associated with the geometry will be executed and afterward the navigator needs to put the particle in the correct geometrical state, depending if it got transmitted or reflected by the surface.
After the propagation, interaction and navigation of the step is done, and the particle has reached its final state for that step, the information about the step and its initial and final particle states can be written to the output.
Subsequently, the active terminators are called to check if the particle has reached a certain physical state where the user has defined it must stop. If this is the case, the track is finished, if not, then the step loop is repeated.

\subsection{Modularity}

The most powerful feature of \textsc{Kassiopeia} is its flexibility and modularity. The user can not only define the type of modules to be used in the simulation such as generators, field calculators or interactions, but the whole composition of the simulation algorithm can be changed depending on the particle's geometrical state.
This is achieved through the combination of the concept of a toolbox in conjuction with a collection of container classes and a dynamic list of commands.
The toolbox exists to maintain all of objects that the user might use during the course of simulation. The container classes (called root classes in \textsc{Kassiopeia}), serve to localize physics processes with similar attributes (discrete interactions, propagation, field calculation, etc.) and execute the activated processes during the main simulation loop. The command list modifies the container classes
and is responsible for the addition, removal or replacement of activated parts of the simulation. It is updated in
a dynamic way depending on the current location of the particle. The details of these three aspects of the simulation are as follows:

\begin{itemize}
	\item \textbf{Toolbox:} All objects of the simulation specified by the user are instantiated and stored in a so called toolbox when the XML configuration file is parsed (see next section \ref{sec:initialization}). This is the case for physics modules such as particle generators, field calculators or interaction processes, but also for completely different objects such as output components (see section \ref{sec:output}).
	\item \textbf{Root classes:} The simulation algorithm works with container classes for the different types of processes displayed in figure \ref{fig:design:flowchart}. These container classes within \textsc{Kassiopeia} are called root classes. There are two different types, divided
	by whether multiple objects  of the same kind are allowed to be active at the same time. The root terminator, for example, can contain a list of multiple terminator objects, so when being called by the simulation algorithm, it will call all its ``child'' terminators. The root trajectory, however, being responsible for the propagation of the particle, can contain only one representation of the equations of motion, although these equations can be composed of multiple terms.	The root classes are typically filled with user-specified default objects at the beginning of the simulation, but they may also be left completely empty. They will then be filled by the navigator depending on the geometrical state of the tracked particle. 
	\item \textbf{Commands:} The manner in which root classes are filled by the navigator is completely exposed to the user. In the configuration file, a single or multiple nested navigation geometries can be defined. For geometric objects (see section \ref{sec:geometry}) a basic distinction is made between a navigation space, a navigation side, (which is some subset of the boundary of a space), and a navigation surface (which is a free surface, not associated with a space). For each navigation geometry, a set of commands can be defined that are executed when the particle enters the corresponding geometry and are reversed, when the particle leaves it. A prerequisite of this command method is that nested geometries must be completely contained within their parent space. This is because processes which are associated with the parent space remain active inside the nested geometry, unless otherwise specified. This helps to avoid unnecessary deactivation and reactivation of objects upon traversal of the geometry. Commands are typically defined to add, remove or replace objects from the root classes. An example of this being the addition of an interaction to the root interaction for a certain surface
	or the replacement of the step size control algorithm for a certain trajectory representation.
\end{itemize}

All objects in the toolbox that are referred to in one of the given commands (and therefore may be used at some time during the simulation) are initialized at the beginning before the start of the simulation. When an object is added to the simulation, it is activated and subsequently deactivated upon removal.
At the beginning of each track, the navigation is started and the simulation is put in a state which depends on the geometry and position of the particle. At each step, the navigator checks if the geometrical state of the particle has changed, and if so, will activate the corresponding navigation geometry and execute the associated commands. After each track is finished, the navigation is stopped and the state of the simulation is put back into the default mode.

With this flexible command concept it is possible to track particles through a variety of different lengths or physical processes with \textsc{Kassiopeia} using a single configuration file. A prime example is the entirety of the \textsc{Katrin} experiment where electrons are generated in a gaseous tritium source, and propagate through the approximately $70\,\mathrm{m}$ long beam line until finally entering a silicon detector. In this case, the dominant physical processes change over the course of a particle's path through each separate region of the experiment. For example, in the source it is important
to consider scattering off tritium molecules, whereas in the UHV of the main spectrometer there is very little scattering, but precision integration
of the equations of motion in the complex electromagnetic fields becomes necessary. An even more drastic change must be executed upon
entry of the silicon material of the detector, since the solid-state physics dictating electron interactions there proceed on a scale of $\mu\mathrm{m}$. This whole journey of the particle can be described as a single track in \textsc{Kassiopeia}, as the simulation algorithm is 
adapted according to the correct underlying physics of each region of the experiment.

\subsection{Initialization}
\label{sec:initialization}

\begin{figure}[tbp]
 %\centering
    \includegraphics{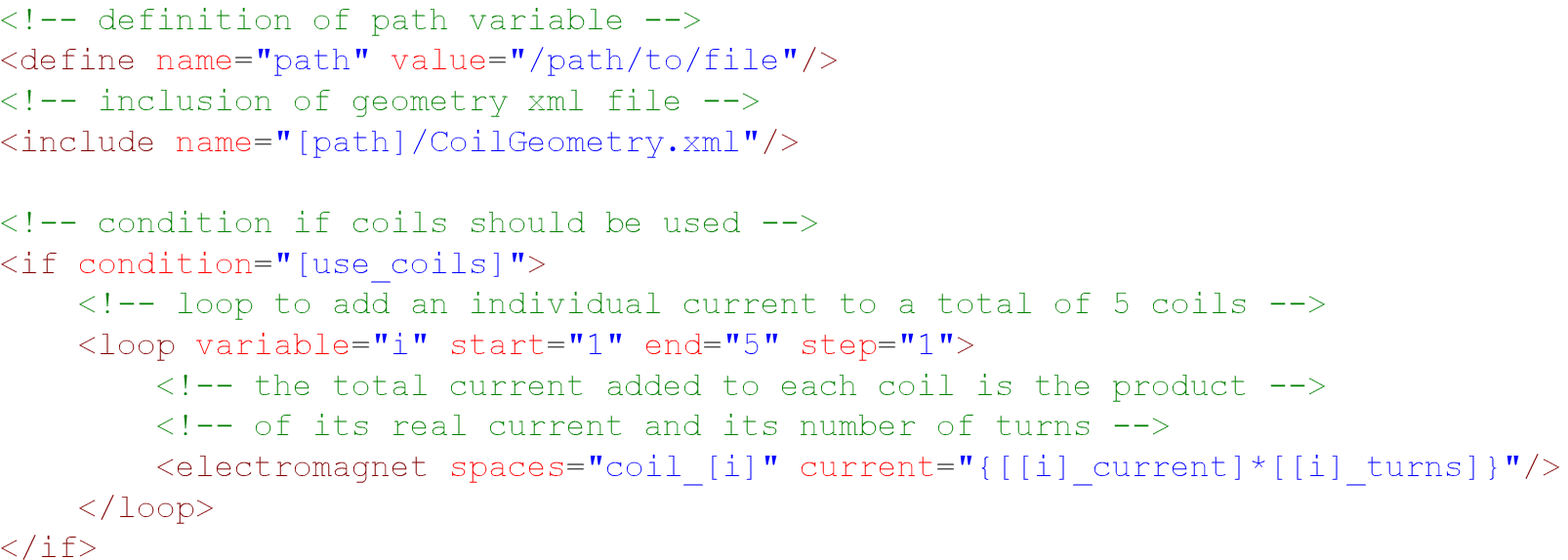}
 \caption[Example of an XML configuration file]{Example of an XML configuration file, showing the features of variables, includes, conditions, loops and formula evaluations.}
\label{fig:initialization}
\end{figure}

A simulation run with \textsc{Kassiopeia} is completely defined by a configuration file. In this file all simulation input data are defined and created. This includes the entire geometry of the experiment, all kinds of different physics processes that may be executed in the simulation, as well as
the level of detail in the output and recorded quantities. The geometry navigation commands and also basic properties of the simulation 
such as the value of the initial random seed and the number of generated events are also defined in the configuration.
The configuration files for \textsc{Kassiopeia} are based on the Extensible Markup Language (XML) as specified in \cite{XMLSpec} with some additional features and extensions. The XML parser of \textsc{Kassiopeia} is composed of a chain of multiple XML processors and the parsing of the information is performed sequentially in the so called SAX style. First an XML tokenizer creates tokens out of the data stream from the file, which are then fed into the chain of XML processors. These processors may modify the stream of tokens until the last processor in the chain finally creates the desired object.
The modifications of the stream by the individual processors allows one the ability to perform advanced operations while the XML file is parsed, such as variable definitions, evaluation of mathematical formulas, or even loops and conditional statements. Furthermore, it is also possible
break the configuration into separate files through an include mechanism or write down a serialized version of a given configuration file.
A snippet of an example configuration file is shown in figure \ref{fig:initialization}.

\section{Geometry}
\label{sec:geometry}

The geometry module of \textsc{Kassiopeia} comprises geometrical classes for a large number of different shapes, linear algebra methods, structures for the relation between geometrical elements and an extension system to add arbitrary properties to shapes.

\subsection{Shapes}

\begin{figure}[tbp]
 %\centering
    \includegraphics{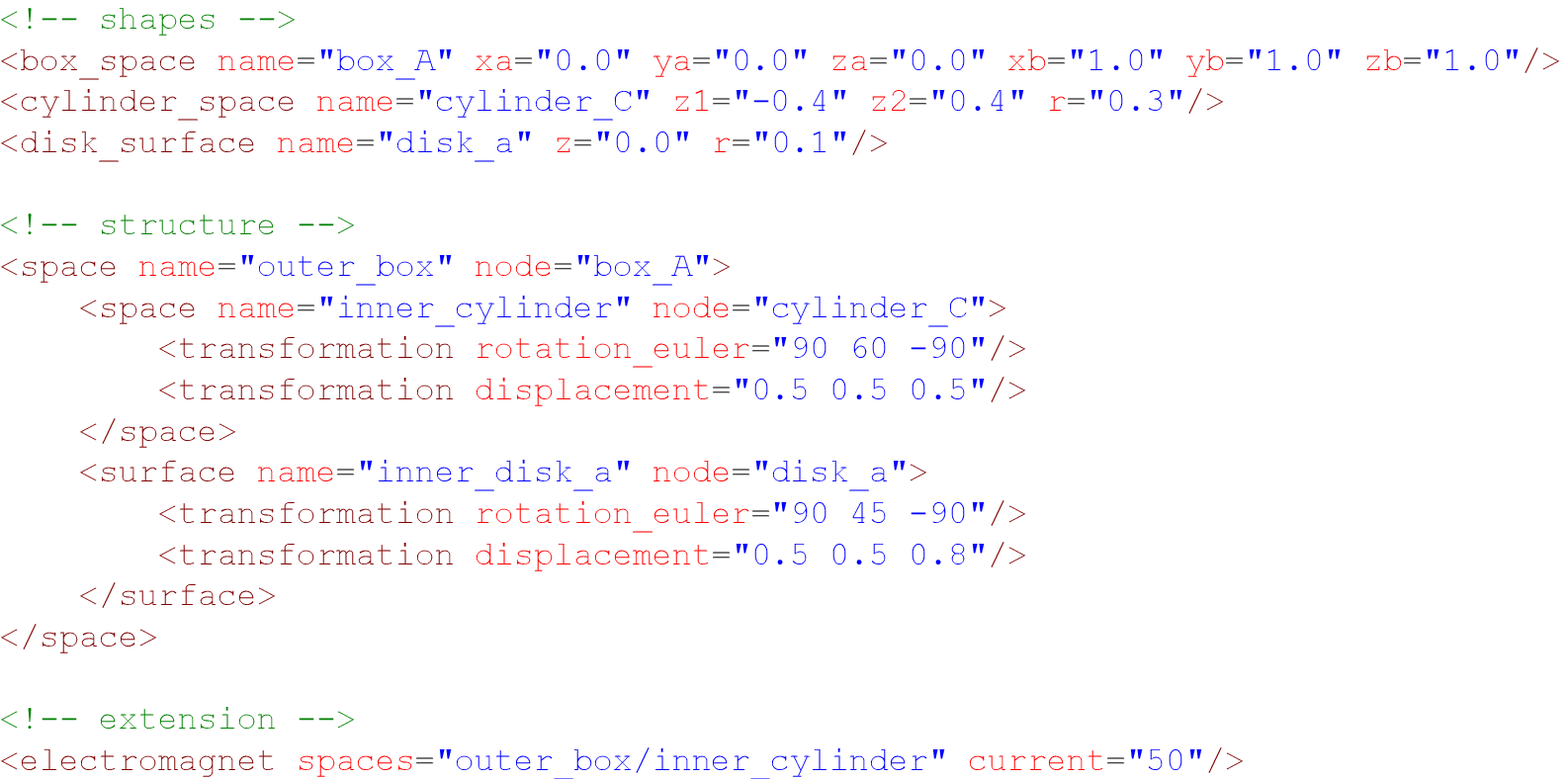}
 \caption[Example of an XML configuration file]{In the geometry XML file, first the definitions of the involved shapes are given, which
 in this example includes a box, a cylinder and a disk surface. Then the relation between the shapes is structured. Here the cylinder and the disk are placed inside the box, and are rotated and displaced in the process. Finally, additional information is added to the cylinder inside the box. In this case, it is made
 into an electromagnet by specifying some associated current. Units of length are in meters, rotation angles are in degrees, and current is given in amperes.}
\label{fig:geometry}
\end{figure}

The available shapes are divided into different types of surfaces and spaces. Both can be constructed from an XML configuration file by defining the necessary attributes that are required for the specific geometrical element. Each single shape is created with its own coordinate system depending on the attribute values chosen by the user.
In example shown in figure \ref{fig:geometry}, a box space, a cylinder space and a disk surface are constructed. The origin of the box is located in one of its corners, while the origin for the cylinder is in its center.

%\lstset{language=XML}
%\begin{lstlisting}
%<box_space name="box_A" xa="0.0" ya="0.0" za="0.0" 
%						xb="1.0" yb="1.0" zb="1.0" />
%<cylinder_space name="cylinder_C" z1="-0.4" z2="0.4" r="0.3"/>
%<disk_surface name="disk_a" z="0.0" r="0.1"/>
%\end{lstlisting}

\begin{figure}[t]
\includegraphics[width = 0.95\columnwidth,keepaspectratio=true]{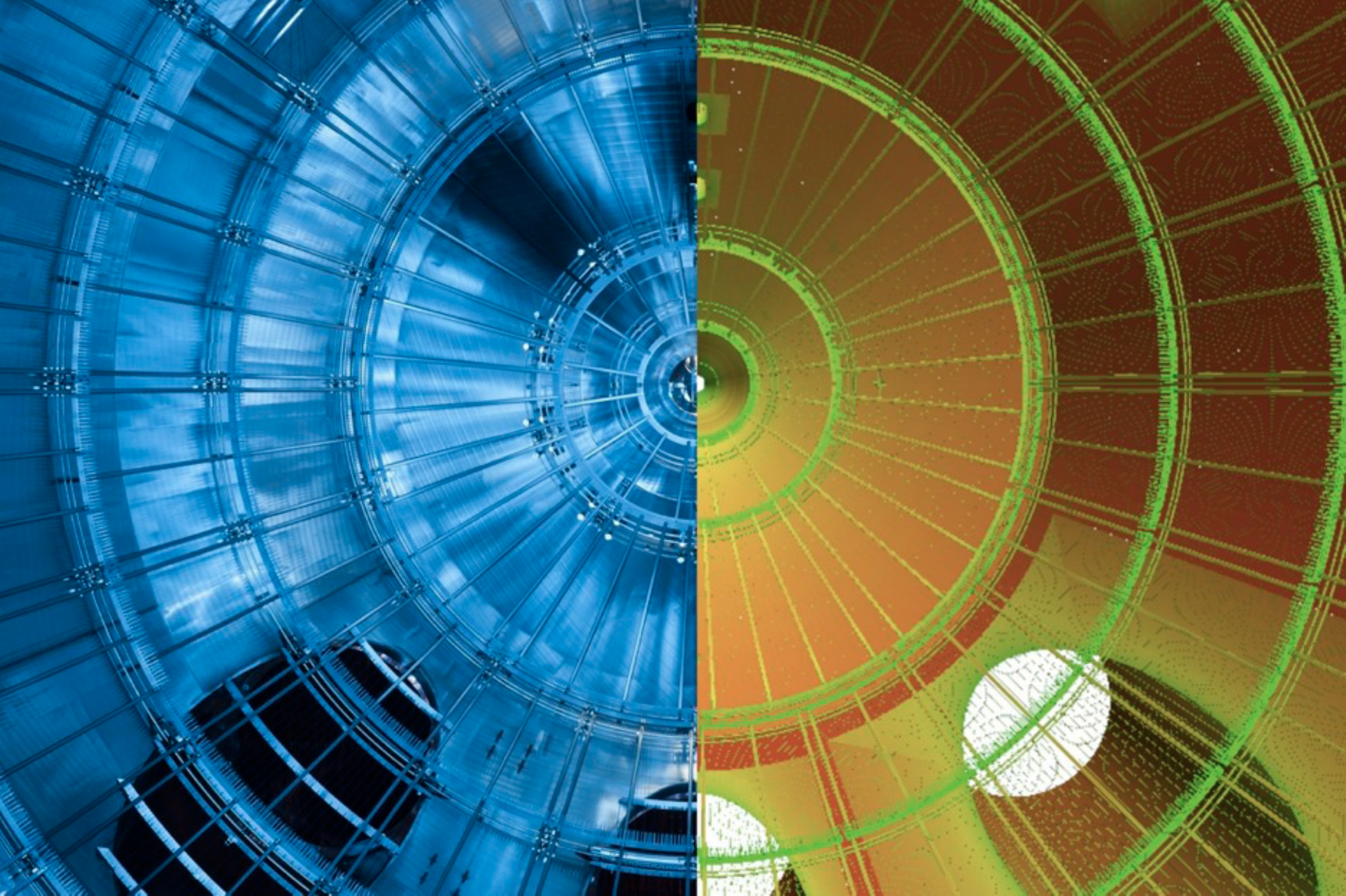}
\caption{A photograph (left side) and 3D electromagnetic model (right side) of KATRIN's main spectrometer.  Visible with both the model and the photograph are the pumping ports, the wire frames and mesh of the inner vessel. }
\label{fig:twosides}
\end{figure}

Aside from the above given examples of basic shapes, it is also possible to construct more complex arbitrary shapes, from a set of points connected either by lines or arcs. The resulting poly-line can be rotated or extruded to create non-trivial surfaces or spaces. All the features of the XML system as described in section \ref{sec:initialization} can be used, such as loops, variable definitions, and mathematical operations. This significantly reduces the amount of text needed to describe a large number of similar geometrical objects. If the available general purpose shapes cannot easily represent
the full complexity of an experiment, special purpose spaces and surfaces can be incorporated by a user following the inheritance mechanism of
the geometry package. As an example of the capability and flexibility of the geometry system, figure \ref{fig:twosides} shows a comparison
between the geometry model and a photograph of the KATRIN main spectrometer interior.

\subsection{Structure}

Surface and spaces have to be placed and related to each other to form a nested relationship. Spaces automatically contain a set of boundary surfaces and may also contain other spaces or surfaces. The policy of \textsc{Kassiopeia}'s geometry module is that child spaces or child surfaces need to be completely contained within their nesting parent space with no protrusion being allowed. The user is responsible for ensuring this condition is satisfied since as of yet no automatic 
collision detection between geometry objects is performed.

\begin{figure}[tbp]
 \centering
		\subfigure[Nested spaces and surfaces]{
      \includegraphics[width=0.45\textwidth]{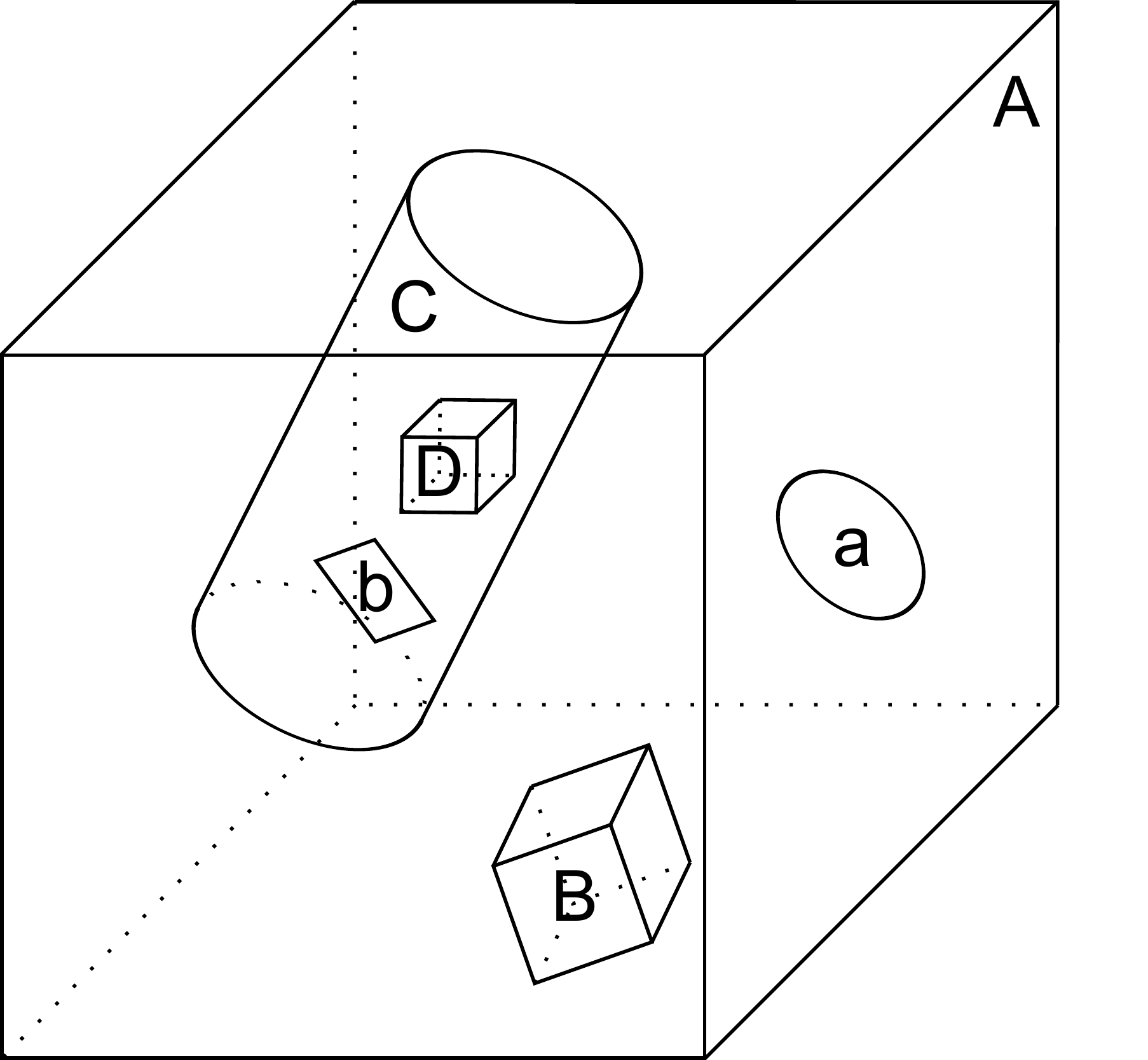}
    }
    \subfigure[Tree structure]{
      \includegraphics[width=0.45\textwidth]{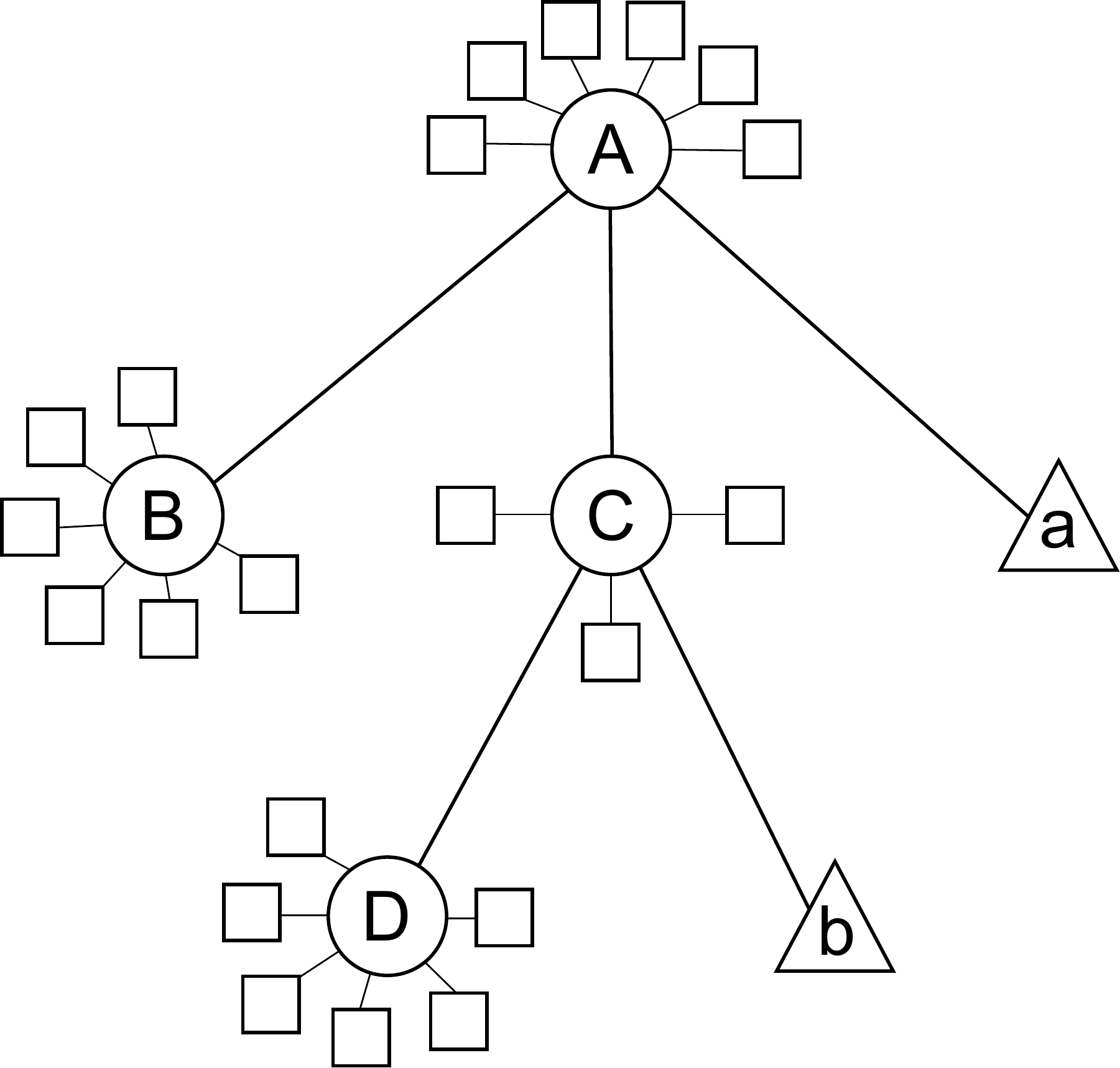}
    }
 \caption[Tree structure of nested geometries]{A set of nested spaces and surfaces (a) and its representation as a tree structure (b). In this example circles represent spaces, squares represent boundary surfaces and triangles represent free surfaces. Figures from \cite{groh:2015}.}
 \label{fig:geometry:tree}
\end{figure}

When a space or surface element is placed inside another space, multiple transformations can be applied to rotate or displace the child space with respect to the coordinate system of the parent space. In the XML example in figure \ref{fig:geometry} the defined shapes are structured in a nested relationship and in figure \ref{fig:geometry:tree} a visualization of this example is drawn together with its corresponding geometry tree. The tree is used to store the geometry relation internally. When referring to a specific space or surface (for example when adding an extension to it), 
its address in the tree is required, this can be specified using an XPath-like syntax as shown in figure \ref{fig:geometry}.

%\lstset{language=XML}
%\begin{lstlisting}
%<space name="outer_box" node="box_A">
%	<space name="inner_cylinder" node="cylinder_C">
%		<transformation rotation_euler="90 60 -90"/>
%		<transformation displacement="0.5 0.5 0.5"/>
%	</space>
%	<surface name="inner_disk_a" node="disk_a">
%		<transformation rotation_euler="90 45 -90"/>
%		<transformation displacement="0.5 0.5 0.8"/>
%	</surface>
%</space>
%\end{lstlisting}

%A visualization of the example with additional elements is drawn in figure \ref{fig:geometry:tree} together with its corresponding geometry tree. The tree is used to store the geometry relation internally. When referring to a specific space or surface, the address in the tree is required, as shown in the next paragraph.

\subsection{Extensions}

The shapes and surfaces of the geometry module feature an extension system which allows a user to append arbitrary information to an object. 
These extensions may contain different types of data for surfaces and spaces. This can be, for the example, the color of a geometrical shape, 
to be used by a visualization module, or an extension with electromagnetic attributes such as the current in a space or the potential on a surface. 
The electromagnetic extensions are of prime importance for the field calculation module in order to compute electromagnetic fields for the given geometry.
In the XML configuration of figure \ref{fig:geometry} an example of an electromagnet is given, where a current of 50A added
as an extension of a cylinder object in order to form a solenoid coil.

%\lstset{language=XML}
%\begin{lstlisting}
%<electromagnet spaces="outer_box/inner_cylinder" current="50"/>
%\end{lstlisting}
\section{Electromagnetic field computations}
\label{sec:fields}
\subsection{Magnetic field}

For the magnetic field calculation of  axisymmetric coils we use the zonal harmonic expansion method
\cite{GlueckZH,Garrett1951}.
This is a special version of the spherical harmonic method, applied for axisymmetric systems.
It can be 100--1000 times faster than the more
widely known elliptic integral method, and  it is more general than the similar, but more widely known, radial series expansion
\cite{paszkowski1968electron,szilagyi2012electron,hawkes1996principles}. 
It has not only high speed
but also high accuracy, which makes the method especially appropriate
for trajectory calculations of charged particles. Due to these properties, no interpolation is necessary
when computing the magnetic field during particle tracking.

The zonal harmonic expansions are convergent at field points within the central and remote
regions, which have spherical boundaries. Their centers, (referred to as the source point), 
can be chosen arbitrarily along the axis of symmetry.
The rate of convergence of the field series depends
 on the distance between the field evaluation point and the source point; the smaller distance for central
field points, and conversely, the larger distance for remote field points, produces a correspondingly higher rate of convergence.
The slower calculation requiring elliptic integrals can be avoided, unless the field point is very close or inside the coil windings.
The zonal harmonic method can also be applicable for general three-dimensional coil systems, as long as the current distribution of
each coil is axially symmetric within its own local coordinate system. See Refs. \cite{GlueckZH}, \cite{GlueckZH2} and \cite{Corona2008} for more details.

The field of non-axisymmetric coils (e.g. coils which produce transverse dipole fields, which are rectangular, rather than solenoidal) are computed
by directly integrating the Biot-Savart formula.
%The field of magnetic materials with known magnetization is 
%computed by equivalent  (fictitious) magnetic charges \cite{Leiber2014}.

\subsection{Electric field}

In KEMField, the boundary element method (BEM) is used for static electric field computations.
In the case of metal electrodes, the (known) boundary conditions given at an
arbitrary point on the electrode
surface may be expressed by a Coulomb integral over the unknown charge density of the
whole surface of the electrode system. Thus an integral equation is obtained for the
charge density function. To handle this problem numerically, 
the surface of the electrodes is discretized into many small boundary elements, and
a system of linear equations is obtained, for which the charge densities of each mesh element serve
as the unknown vector. To solve this system of equations, either a direct or an iterative method is used.
Once the charge densities are known, the potential and field at an arbitrary point can then be
computed by summing the potential and field contributions over all boundary elements.
In KEMField, both  metal surfaces (with Dirichlet boundary conditions)  and also dielectrics 
(with Neumann boundary conditions) can be used for electric field computations. \cite{phd:corona}.

The BEM has several advantages relative to the finite difference method (FDM) and 
the finite element method (FEM). Firstly, there is no need to discretize the whole
three-dimensional volume, only the two-dimensional surfaces, thus the number of
degrees-of-freedom is usually smaller. Secondly, for a given 
computation time, the BEM produces more accurate potential and field values
than FDM and FEM \cite{Cubric}. In addition, with BEM, the potential and field at arbitrary points
can be computed directly from the charged sources (rather than interpolated from values known at fixed points),
which leads to very high accuracy and yields smooth field solutions. BEM is also ideal 
for an electrode system with large differences in size scales (e.g. very small electrodes in a large volume vessel)
and can easily handle open systems, in contrast to FDM and FEM.

In the special case of applying the BEM for axially symmetric electrodes ,
the potential-field contribution of a boundary element is usually evaluated with the help of the first and
second complete elliptic integrals; often, a numerical integration of the elliptic integral formulas may also be
necessary. To compute the potential and field of the whole electrode system, these contributions have to be
summed over all elements. This summation can be rather time-consuming. Fortunately, 
it is possible to use the previously mentioned zonal harmonic expansion method
within a large region of space of an axisymmetric electrode system \cite{GlueckZH3},
and this method  is 2-3 orders of magnitude faster than the calculation using elliptic integrals.

In the case of axisymmetric electrode system, a few hundred or few thousand elements are usually enough to
get an appropriate discretization. A direct method, like Gauss-Jordan elimination or LU decomposition, can be used 
in that case to compute the unknown charge densities of the elements. 
Direct methods can also be used for the special case of a discrete rotationally symmetric electrode system,
where the number of elements could be high (e.g. few millions) but the number of different charge densities
is still small (e.g. a few thousand).

In the more general case of a complicated three-dimensional electrode system, a few hundred thousand or
few million elements might be necessary for a good discretization of the original surfaces. Typically, a three-dimensional electrode system 
is discretized by a combination of triangles, rectangles or wire elements. These basic geometric elements approximate the charge density on 
the modeled surface using a piecewise constant function for the charge density. With large linear systems of this size, 
memory requirements render direct methods inapplicable and iterative methods become necessary. 
KEMField deals with large systems through the use of either the Robin Hood method, or Krylov subspace methods.
The Robin Hood method, which is a special version of the Gauss-Seidel iteration, allows one to solve the
linear system with a memory cost proportional to $\mathcal{O}(N)$ and a computational cost which scales
like $\mathcal{O}(N^\alpha)$ (with $1 < \alpha < 2$) \cite{RobinHood}. On the other hand, 
Krylov subspace methods such as GMRES \cite{saad1986gmres}, if used with straight-forward matrix-vector
multiplication, would by themselves generally be insufficient to efficiently solve problems of this size. 
However, when combined with fast multipole techniques \cite{grengard1988rapid,rokhlin1985rapid} they are a powerful tool
which can greatly reduce the time to solution and reduce memory cost. While a full description of the specific algorithm used 
by the KEMField library is beyond the scope of this paper, KEMField provides a modified multipole method
which is a hybrid of the traditional Fast Multipole Method (FMM) \cite{greengard1997new} and a Fourier transform based 
approach known as the Fast Fourier Transform on Multipoles (FFTM) \cite{ong2003fast,lim2008fast}, which is described in detail in \cite{barrett:2016}. Krylov subspace
methods benefit greatly from preconditioning when dealing with the three dimensional Laplace BEM and KEMField provides
several simple choices such as Jacobi and Block-Jacobi, as well as an implicit preconditioner which acts by solving the same
problem at reduced accuracy at each iteration in order to very effectively reduce the number of full accuracy iterations needed.

Once the charge densities are known, the potential and field of an element can then be computed directly though the use of either analytical or
numerical integrations. Typically, analytical integration formulas are used to deal with the singular integrals when the evaluation point 
is close to the element, while Gaussian quadrature which exhibits greater numerical stability is used for field points which are far 
from the element (relative to the element size). Direct calculation of the potential and field by summing over all elements during 
charged particle tracking is very time consuming. To reduce the computation time, one method that is provided is 
an interpolating field map grid. For a fixed electrode configuration, the potential and field values at the 
many grid points have to be evaluated only once, and during charge particle  tracking the field calculation by 
interpolation is much faster than by the direct summation method. In order to increase the accuracy and to 
reduce the memory requirements, we use a cubic Hermite interpolation procedure \cite{Eupper,Corona2008,Leiber2014}.
This method is very effective when particle tracking is performed within small volumes which are 
well removed from boundaries, and fast and accurate field evaluation is needed. 
Another method which can sometimes require greater memory usage but can map the field to within a user defined 
tolerance everywhere is provided by the Fast Multipole Method. This technique constructs a large collection of spherical
multipole expansions (of the boundary-elements in the far-field) covering the volume of interest, 
while near-field boundary elements have their field contributions evaluated directly. Several parameters
exposed by this method, such as the expansion degree and spatial resolution allow the user great flexibility to fine tune
a compromise between the accuracy, memory usage, and the speed of field evaluation at run time.

In order to make full exploitation of modern computing resources, KEMField has been designed to take advantage of 
parallelization whenever possible. In order to do this KEMField can make use of Graphics Processing Units (GPUs) 
using OpenCL \cite{stone2010opencl} to accelerate both field calculation as well as solving of BEM problems. Additionally,
KEMField can be compiled with MPI \cite{gabriel2004open}, for use on distributed computing platforms making use of either 
CPU or CPU+GPU architectures.

\section{Generation of particles}
\label{sec:generation}

At the beginning of each event a particle generator needs to produce the initial state for a set of particles. Besides the definition of a particle's intrinsic nature via its mass and charge, it needs to be fully characterized by seven parameters: position ($x, y, z$), momentum ($p_x, p_y, p_z$) and time ($t$). While the type of the particle and therefore its inherited properties can easily be specified by an ID number, following the PDG particle numbering pattern \cite{pdg}, the generation of its dynamic properties is broken up into a independent substructure consisting of four generators for the particle properties: time, position, energy and direction. This choice of quantities was motivated by the particular use case of KATRIN, which needs to
examine particle motion in a MAC-E filter for which it is generally advantageous to initialize a 
particle state by setting its energy and momentum direction, rather than setting the momentum vector directly. 

For each of the four independent quantities (time, position, energy and direction) the generated values can be set independently
by specifying so-called value generators, which can drawn numbers from a user specified distribution. 
Any combination of these value generators can be used to initialize the four basic quantities, leading to a large number of possibilities.
A selection of value generators that are available within \textsc{Kassiopeia} is presented in table \ref{tab:generation:value}.

\begin{table}
  \centering
  \caption{Selection of Value Generators with description}
  \begin{tabular}[tbp]{lp{.7\textwidth}}
		\toprule
		Value Generator	& Description \\
		\midrule
    Fix  					&  A fixed value defined by the user.	\\
		Uniform		 		&  The value is drawn from a uniform distribution between a defined minimum and maximum value.	\\
		Gauss					&  The value is drawn from a Gaussian distribution with a mean $\mu$ and standard deviation $\sigma$. Generated values may optionally be limited between a defined minimum and maximum value with the normalization adjusted accordingly. 	\\
		Formula				&  The value is drawn between a minimum and maximum value according to a density distribution of a user-defined formula.	\\
		Set						&  A specific number of values is generated equally spaced between a start and an end value.	\\
		List					&  A list of values is used where each value in the list can be specified by the user.	\\
		Cylindrical		&  A position value is drawn uniformly from a cylindrical volume.	\\
		Spherical			&  A position or direction value is drawn uniformly from the surface of a sphere.	\\
    \bottomrule
  \end{tabular}
  \label{tab:generation:value}
\end{table}

These value generators can be used separately or combined to form a composite generator
for time, position, energy and direction. In addition, 
there exist some special generators, which do not make use of the composite value mechanism. 
For example, these can be special position generators which can create random initial positions
uniformly in a space or over a surface, or energy generators which reflect the
radioactive decay sequences of specific unstable isotopes of radon, krypton or lead.

\section{Propagation of particles}
\label{sec:propagation}

The propagation of particles, or more specifically, the calculation of their corresponding trajectories is one of the most important parts
in the \textsc{Kassiopeia} software. This module is responsible for integrating the equations of motion which are represented as first order ordinary differential equations. Within \textsc{Kassiopeia} all continuous physics processes are represented as terms in the overall equation of motion. This includes not only the propagation of the particle in conservative force fields, but also processes such as radiative synchrotron losses, which can separately be included in the equation of motion. In this way, when numerically solving the particle motion, all of the included terms can treated together on equal footing.
Depending on the choice of variables used for the full physical state of a particle, the terms present the equations of motion can adopt different representations. These are referred to as ``trajectories'' in \textsc{Kassiopeia}. Each of the available trajectory types will be introduced in the following subsections, along with the specific differential equations for each physics process. We will also outline the integrators used 
to solve the differential equations in conjunction with the step size controls required to mitigate numerical error.

\subsection{Exact Trajectory}

When dealing with an exact trajectory, the physical state of the propagating particle is described by a pair of variables which are implicitly
a function of the time coordinate $t$. These are the position vector, $\mathbf{r}$, and the momentum vector, $\mathbf{p}$.
The representation of the propagation term in the equation of motions of a particle with charge $q$ in an electric and magnetic field $\mathbf{E}$ and $\mathbf{B}$ is given by the Lorentz equation. The terms for the ordinary differential equations for the variables of the exact representation are therefore:
\begin{eqnarray}
\frac{\mathrm{d}\mathbf{r}}{\mathrm{d}t} & = & \frac{\mathbf{p}}{\gamma m}\;, \\
\frac{\mathrm{d}\mathbf{p}}{\mathrm{d}t} & = & q \left( \mathbf{E} + \frac{\mathbf{p} \times \mathbf{B}}{\gamma m} \right) \;,
\label{eq:propagation:exact}
\end{eqnarray}
where $m$ is the rest mass of the particle, and $\gamma$ is the relativistic Lorentz factor. 

\subsection{Adiabatic Trajectory}

\begin{figure}[tbp]
 \centering
		\subfigure[Exact trajectory]{
      \includegraphics[width=0.47\textwidth]{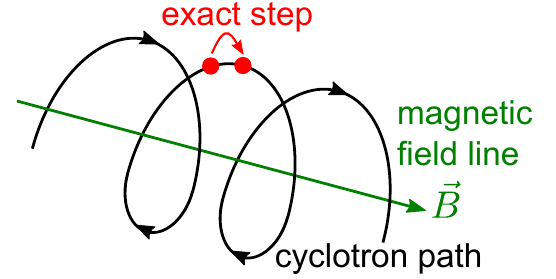}
    }
    \subfigure[Adiabatic trajectory]{
      \includegraphics[width=0.47\textwidth]{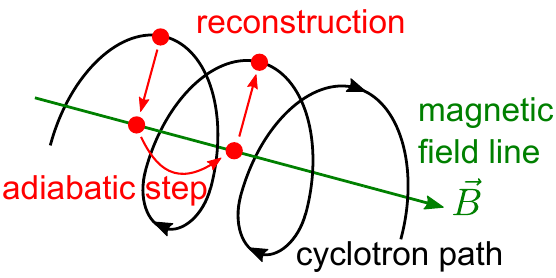}
    }
 \caption[Illustration of the exact and adiabatic trajectory]{Illustration of the step taken when computing the exact trajectory (a) and the adiabatic trajectory (b). In the adiabatic trajectory calculation the guiding center position is propagated, which allows a much larger step size. The particle's exact position 
 can be reconstructed afterwards if the gyration is included. Figures from \cite{groh:2015}.}
 \label{fig:propagation:stepping}
\end{figure}

If the magnetic and electric fields are nearly constant within a cyclotron radius of the particle, the first adiabatic invariant $\gamma \mu$, with $\mu$ being the magnetic moment, remains conserved along the trajectory of the particle. Under this approximation, the physical state of the particle can be represented by its time, $t$, the guiding center of the motion, $\mathbf{r}_\mathrm{c}$, the components of the particle's momentum which are parallel and perpendicular to the magnetic field $p_\parallel$ and $p_\perp$, and the gyration phase $\phi$ \cite{northrop1961guiding}. The advantage of using this adiabatic trajectory is the much larger step size that is possible compared to the case of calculating an exact trajectory, since the curvature in the path of the guiding center position is much smaller than in the propagation of the real particle position. The exact position of the particle can be reconstructed from the guiding center after the propagation step, as visualized in figure \ref{fig:propagation:stepping}.

In case of the adiabatic representation, the propagation terms in the ordinary differential equation assume the following form:
\begin{eqnarray}
\frac{\mathrm{d}\mathbf{r}_{c}}{\mathrm{d}t} & = & \frac{p_{\parallel}}{m\gamma} \frac{\mathbf{B}_\mathrm{c}}{B_\mathrm{c}}, \\
\frac{\mathrm{d}{p}_{\parallel}}{\mathrm{d}t} & = & - \frac{ p_{\perp}^{2} }{2 \gamma m B_\mathrm{c}} \mathbf{\nabla} B_\mathrm{c} + q \mathbf{E}_\mathrm{c} \cdot \frac{\mathbf{B}_\mathrm{c}}{B_\mathrm{c}}, \\
\frac{\mathrm{d}{p}_{\perp}}{\mathrm{d}t} & = & \frac{p_{\perp} p_{\parallel}}{2 \gamma m B_\mathrm{c}} \mathbf{\nabla} B_\mathrm{c} \cdot \frac{\mathbf{B}_\mathrm{c}}{B_\mathrm{c}} \;.
 \label{eq:propagation:adiabatic}
\end{eqnarray}
Two additional terms need to be added to these propagation terms to account for gyration and drift caused by the magnetron motion.
The gyration term can be derived from the cyclotron frequency of the particle and since it completely decouples from the guiding
center motion is simply:
\begin{equation}
\frac{\mathrm{d}\phi}{\mathrm{d}t} = \frac{q B_\mathrm{c}}{m\gamma},
 \label{eq:propagation:adiabatic:gyration}
\end{equation}
The terms accounting for the drift motion are more complicated as they modify both guiding center position as well as the momentum components,
their contributions can be written as:
\begin{eqnarray}
\left. \frac{\mathrm{d}\mathbf{r}_\mathrm{c}}{\mathrm{d}t} \right |_{\mathrm{drift}} & = & \frac{\mathbf{E}_\mathrm{c} \times \mathbf{B}_\mathrm{c}}{B_\mathrm{c}^2} + \frac{2 p_\parallel^2 + p_\perp^2}{qm(\gamma + 1)B^3_\mathrm{c}} \mathbf{B}_\mathrm{c} \times \mathbf{\nabla} B_\mathrm{c} \frac{\mathbf{B}_\mathrm{c}}{B_\mathrm{c}},  \\
\left. \frac{\mathrm{d}{p}_{\parallel}}{\mathrm{d}t} \right |_{\mathrm{drift}} & = & \frac{q\gamma m}{p_\parallel} \mathbf{E}_\mathrm{c} \cdot \frac{\mathrm{d}\mathbf{r}_\mathrm{c}}{\mathrm{d}t} - \frac{p_\perp^2}{2p_\parallel B_\mathrm{c}} \left( \mathbf{\nabla} \mathbf{B}_\mathrm{c} \cdot \frac{\mathrm{d}\mathbf{r}_\mathrm{c}}{\mathrm{d}t} \right) \cdot \frac{\mathbf{B}_\mathrm{c}}{B_\mathrm{c}}, \\
\left. \frac{\mathrm{d}{p}_{\perp}}{\mathrm{d}t} \right |_{\mathrm{drift}} & = & \frac{p_\perp}{2B_\mathrm{c}} \left( \mathbf{\nabla} \mathbf{B}_\mathrm{c} \cdot \frac{\mathrm{d}\mathbf{r}_\mathrm{c}}{\mathrm{d}t} \right) \cdot \frac{\mathbf{B}_\mathrm{c}}{B_\mathrm{c}} \;.
 \label{eq:propagation:adiabatic:drift}
\end{eqnarray}

\subsection{Magnetic Trajectory}

An additional trajectory type within \textsc{Kassiopeia} is the ``magnetic trajectory'' which can be used to visualize magnetic field lines. This is achieved by creating a pseudo particle which is represented only by its time $t$ and position $\mathbf{r}$. Particle properties such as kinetic energy or momentum are undefined.
In this regard, it is interesting to note that in the context of KATRIN, electrons with a kinetic energy in the keV regime posses relatively small cyclotron radii. Therefore, on occasion, a magnetic trajectory can be a good approximation for the path of such an electron. This simplification can be advantageous since
a magnetic trajectory can be calculated with considerably less computational effort than that which is required to track a charged particle either exactly or adiabatically.
The ordinary differential equation in this case is simply:
\begin{eqnarray}
\frac{\mathrm{d}\mathbf{r}}{\mathrm{d}t} & = & \frac{\mathbf{B}}{B}\;.
\label{eq:propagation:magnetic}
\end{eqnarray}

\subsection{Synchrotron losses}

Besides the previously mentioned terms for the propagation of the particle, other terms can be added as well, such as the synchrotron term, which handles free-space radiative losses by accelerated particles.  
This radiated power needs to be expressed as a force term for use in trajectory calculations. The well known expression for this is the Abraham-Lorentz-Dirac radiation reaction force which suffers from several pathologies such as violation of energy conservation and causality, as pointed about by Dirac \cite{Dirac:1938}. Within \textsc{Kassiopeia} a comparatively new approach is implemented that avoids these problems and follows the equation of motion proposed by Ford
and O'Connell \cite{FordConell:1993}, which reads as:
\begin{equation}
	m\frac{d(\gamma \mathbf{v})}{dt} = \mathbf{F} + \tau_e\left[\gamma\frac{d\mathbf{F}}{dt}-\frac{\gamma^3}{c^2}\left(\frac{d\mathbf{v}}{dt}\times\left(\mathbf{v}\times \mathbf{F}\right)\right)\right],
\end{equation} 
where $\tau_e=2e^2/3m_ec^3$ and $\mathbf{F}$ is the known Lorentz force:
\begin{equation}
	\mathbf{F}=q\left(\mathbf{E}+\mathbf{v}\times\mathbf{B}\right),
\end{equation}
while the second term is the radiation reaction force. This is the force term that is used in \textsc{Kassiopeia} for the exact trajectory. Notice that the total time derivative of the fields requires us to calculate the field gradient and these are in general very costly to evaluate.
However, the term accounting for radiative losses in the adiabatic representation is simpler and faster. 
The full derivation of which can be found in \cite{furse:2015}. 
The radiated power can be described in terms of external forces acting on the particle as follows:
\begin{equation} 
P = \frac{\mu_{0}}{6 \pi c} \frac{q^{2}}{m^{2}} \left( f_{\parallel}^{2} + \gamma^{2} f_{\perp}^{2} \right),
\label{eq:propagation:sync:power_forces}
\end{equation}
where $f_{\parallel}$ is the component of the force acting tangential to the particle's motion and $f_{\perp}$ acts normal to its motion.
Equation \ref{eq:propagation:sync:power_forces} needs to treated approximately by making the assumption that the motion of the gyration
is completely responsible for the radiation. This assumption is completely consistent with the adiabatic approximation, 
and yields the complete synchrotron term in this representation:
\begin{eqnarray}
\left. \frac{\mathrm{d}\mathbf{r}_{c}}{\mathrm{d}t} \right |_{\mathrm{sync}} & = & 0  \;,\\
\left. \frac{\mathrm{d}\phi_{c}}{\mathrm{d}t} \right |_{\mathrm{sync}} & = & 0  \;,\\
\left. \frac{\mathrm{d}{p}_{\perp}}{\mathrm{d}t} \right |_{\mathrm{sync}} & = & - \frac{\mu_{0}}{6 \pi c} \frac{q^{4}}{m^{3}} {\left| \mathbf{B}(\mathbf{r}_{c},t) \right|}^{2} {p}_{\perp} \gamma \;, \\
\left. \frac{\mathrm{d}{p}_{\parallel}}{\mathrm{d}t} \right |_{\mathrm{sync}} & = & 0 \; .
\label{eq:propagation:sync:adiabtic}
\end{eqnarray}
This description is free of field gradients and as such can only be an approximation. However, the adiabatic approximation to hold, the gradients need to be negligible anyway and therefore no new assumption is made here.

\subsection{Integrators}

All of the mentioned trajectories can be written as ordinary differential equations of the form $y' = f(x,y)$.
These differential equations have to be integrated numerically. The numerical integration is carried out
by integrators of the Runge-Kutta type. \textsc{Kassiopeia} provides a variety of Runge-Kutta integrators which may be chosen
by the user depending on their preference for accuracy or speed. Table \ref{tab:trajectory:integrators} lists the available integrator types and some
of their properties. These integrators can be categorized by several properties. These are; the order of truncation error on the solution as 
a function of the step size, the number of function evaluations required to propagate the solution at each step, 
and the availability of a local error estimate. Since straightforward Runge-Kutta routines typically do not provide information 
about the solution in between steps, some integrators have been equipped with a continuous interpolant (dense output).  
All Runge-Kutta integrator routines which do not provide an embedded dense output can optionally be equipped with a first or third 
order Hermite interpolant for dense output which requires no additional function evaluations. For problems where extremely 
long tracking times or excellent energy conservation is required, a symplectic integrator is also provided, as detailed in the table \ref{tab:trajectory:integrators}. Since \textsc{Kassiopeia} is object 
oriented and easily extensible, users may also chose to add their own integrators using alternate Runge-Kutta, predictor-corrector
or symplectic algorithms. Any additional integrators can be accomodated as long as they inherit from the KSMathIntegrator interface.

\begin{center}
\begin{table}[h!]
  \caption{List of supplied Runge-Kutta integrators}
    \begin{tabular}{ p{1.7cm} p{1.4cm} p{2.2cm} p{2cm} p{2.5cm} p{2cm}}
		\toprule
    Name    & Solution order & Function evaluations per step & Embedded local error estimate & Dense output order & Reference \\
    \midrule
    RKF54   & 5              & 6                             & Yes                           & NA                 & \cite{fehlberg1969low} \\
    RKDP54  & 4              & 7                             & Yes                           & 5                  & \cite{dormand1980family,hairer1993solving} \\
    RK65    & 6              & 8                             & Yes                           & NA                 & \cite{prince1981high}  \\
    RK8     & 8              & 13                            & No                            & NA                 & \cite{verner1978explicit} \\
    RK86    & 8              & 12                            & Yes                           & NA                 & \cite{tsitouras1999cheap} \\
    RK87    & 8              & 13                            & Yes                           & NA                 & \cite{prince1981high} \\
    RKDP853 & 8              & 16                            & Yes                           & 7                  & \cite{hairer1993solving} \\
    SYM4\footnotemark   & 4              & 4                             & No                            & NA                 & \cite{chin2008symplectic} \\
    \bottomrule
		\end{tabular}
    \label{tab:trajectory:integrators}
\end{table} 
\end{center}
\footnotetext{This integrator can only be used with the exact trajectory.}

\subsection{Controls}

The computation of a step is critical in terms of performance. Since it usually requires the calculation
of the electromagnetic field at multiple positions. Therefore the number of calculated steps for each track should be limited as much as possible.
This must be done carefully, as the accuracy in calculating the particle's motion will decrease for larger step sizes, which can lead to a violation of energy conservation.
The step size of the particle therefore needs to be adjusted to meet the specific requirements of the user. This customization is handled by step size controls, which are part of the tracking module of \textsc{Kassiopeia}.
A step size control suggests a specific step size (with the dimension of time) to the integrator at the beginning of each step. 
After the step has been performed the step size control may accept or reject the current step, and suggest a new step size. 
It is also possible to use multiple step size controls simultaneously with the smallest suggestion being used.
The simplest step size controls within \textsc{Kassiopeia} are user defined constraints on maximum time step or a maximum step length.
More advanced step size controls are also available. For regions where a magnetic field is present the step size 
can be chosen based on a fixed fraction of the current cyclotron period of the particle, this control results in small steps in high magnetic fields and large steps in low ones. 
It also possible to adjust the step size dynamically to keep the violation of conserved quantities, such as the total energy, within a user-defined range. Additionally, dynamic step
size adjustment is also available if a Runge-Kutta integrator with embedded error estimation is used. In this case step size can
be adjusted to keep the local error on the position and/or momentum of the particle within a range of user defined absolute or relative tolerances.

\section{Interaction of particles}
\label{sec:interaction}

The interaction processes within \textsc{Kassiopeia} are grouped into space interactions and surface interactions.
Space interactions occur stochastically as a function of a given probability during particle propagation within a dense medium, while surface interactions are triggered only when the particle reaches a surface which has a specific interaction attached to it.

\subsection{Space interactions}

For particles moving with a velocity $v$ through a medium with a target number density of $n$, the probability for an interaction with cross section of $\sigma$ to occur after time $t$ can be calculated according to:
\begin{equation}
P(t) = 1 - \exp\left(-\frac{t \cdot v}{\lambda}\right)\;,
\label{eq:interaction:probability}
\end{equation}
where the mean free path is defined as
\begin{equation}
\lambda = \frac{1}{n \cdot \sigma}\;.
\label{eq:interaction:mean_free_path}
\end{equation}
In \textsc{Kassiopeia} the parameters $n$, $v$ and $\sigma$ are calculated as mean values for the initial and final position of the particle on the step.
The density of the medium is calculated by a separate density module. The simplest example being defined by a constant density, which may
be related to properties specified by user such as partial gas pressure and temperature.
The cross section $\sigma$ corresponds to the sum of all individual cross sections of interaction processes for this scattering module.
The time for the next scattering to take place can be generated stochastically according to:
\begin{equation}
t_\mathrm{scat} = - \ln\left(1 - P\right) \cdot \frac{\lambda}{v}\;,
\label{eq:interaction:interactiontime}
\end{equation}
where $P$ is drawn from a uniform distribution between 0 and 1.  If this scattering time is larger than the time it takes the particle to complete the current step, no scattering will take place. On the other hand, if $t_\mathrm{scat} < t_\mathrm{step}$, the trajectory of the particle is recalculated with a step size identical to the scattering time so that the properties of the particle are exactly computed at the time just before scattering takes place. After that, a decision on the specific scattering process to be executed (typically elastic or inelastic scattering), is made based on their individual cross section contributions. Finally, the scattering process is executed, thereby modifying the properties of the particle and optionally creating new particles.% for an ionization process.

\subsection{Scattering types}

Scattering processes are treated by the simulation in a modular way so that each process can be handled individually as a so-called scattering calculator. This unit is responsible for calculating the cross section as a function of the particle's state, as well as executing the interaction process
which modifies the particle. Multiple calculators with the same species can then be grouped into a scattering module.

The dominant gas species in \textsc{KATRIN}'s main spectrometer is molecular hydrogen, while in the source region it is molecular tritium, with the scattering properties of both isotopes being very similar.
The corresponding cross sections, energy losses and angular changes for elastic \cite{scattering_diff_xsection}, \cite{ElasticHydrogen1}, \cite{ElasticHydrogen2}, excitation \cite{ExcitationHydrogen}, \cite{ExcitationHydrogen2} and ionization processes \cite{Ionisation} are implemented within different scattering calculators, one for each process.

Additional scattering processes for different species can be implemented easily. For example, this can be done by importing data from the LXcat database \cite{lxcat}, which has already been accomplished for the process of electron scattering off argon \cite{lxcatargon}.

In addition, specific scattering calculators to describing electron interactions in silicon are available. These were adapted from the \textsc{Kess} package \cite{phd:renschler} and now are completely integrated into \textsc{Kassiopeia}.

%Additionally, a scattering calculator is available with a fixed cross section and the energy loss and angular change is diced according to a user defined function. This is need for example for the scattering in KATRIN's gaseous source, as it will be detailed in chapter \ref{ch:simulation}.

\subsection{Surface Interactions}

Surface interactions differ from space interactions as they are only enforced when the particle crosses a specific surface. The associated surface interaction has to decide whether the particle is transmitted to the next space or reflected back into the previously occupied space. This will modify the properties of the particle accordingly, and can result in an angular or energy change when crossing the interface between different materials. Of particular importance to 
KATRIN are processes where particles enter from vacuum into a solid silicon detector.
\section{Termination of particles}
\label{sec::termination}

The trajectory of a particle has to be terminated once a specific condition from a user-defined set is met. The condition to terminate the propagation of the actual particle can be defined in a very flexible way. For example, the termination condition can be determined to occur when the particle hits a detector after having propagated through an experiment, or it may be made by identifying a specific particle property that makes further tracking meaningless, such as when a certain minimal kinetic energy is reached, or if an undesired propagation direction of the particle manifests. A selection of terminators available within \textsc{Kassiopeia} are presented in table \ref{tab:termination}.

\begin{table}
  \centering
  \caption{Selection of Terminators with description}
  \begin{tabular}[tbp]{lp{.7\textwidth}}
		\toprule
		Terminator Name	& Description \\
		\midrule
    min/max z  							& upper/lower bound on z component of position	\\
    min/max r  							& upper/lower bound on r component of position	\\
		min/max kinetic energy 	& upper/lower bound on kinetic energy \\
		max time	 							& upper bound on track time	\\
		max length							& upper bound on track length	\\
		max steps								& upper bound on track step count	\\
		death										& stops track if this terminator is active	\\
		secondaries							& stops track if particle is a secondary	\\
		min distance						& lower bound on distance to specific space or surface	\\
		trapped									& upper bound on longitudinal momentum sign change count	\\
		output									& allowed range for the value of a specific output variable	\\
    \bottomrule
  \end{tabular}
  \label{tab:termination}
\end{table}

As a terminator module is a small class with a simple structure, additional terminators as required by the user can easily be added.
Like all other modules, terminators can be attached to specific navigation spaces or surfaces of the simulation and will only be activated once the particle enters that specified geometry.
\section{Output}
\label{sec:output}

The output that will be written to disk for a given simulation depends on the users needs. A static output system that writes down fixed particle properties after each track or even step is therefore not desirable. This becomes evident when running Monte Carlo simulations with millions of tracks, each track containing millions of steps, as the disk space required to save the simulated information can easily reach problematic dimensions.

Therefore, the output system for the recent version of \textsc{Kassiopeia} was designed in a highly flexible way that allows the user to define each individual output component in the XML configuration for the four levels of detail: run, event, track and step. Therefore, it is possible to store exactly the amount of information that is required, and the output can be made to include very specific information which is desired from relevant objects in the simulation.

%In the following the working principle of the output components will be introduced. Then the analysis logic that can be applied to the output stream while writing to disk will be presented, which is followed by a description of the available output writers and readers.

\subsection{Output components}

A single output component can be configured through a chain-like system which starts with an object that has been put into the simulation toolbox. 
By calling a getter function of the simulation object and repeating the procedure for the resulting object, a chain can be produced with the desired output variable at the end.
For example, if a user wants to write down the magnitude of the magnetic field at the end of each step. This can be done by retrieving the step object from the toolbox, getting the final particle object from the step, getting the magnetic field vector from the particle and finally, calculating the magnitude of the magnetic field vector.

Multiple output components with the desired information about the simulation need to be grouped together before they can be added to the object writing the output. Each group corresponds to a tree in the \textsc{Root} \cite{root} data structure, which is written to disk en bloc. When adding an output group to the writer object, the user can choose the level at which this group should be written, i.e.\ each step, track, event or run. Output groups at the step level can be connected to the navigation geometry as well as any other simulation module, and will only be turned-on when the particle is in an active region. A typical step output group might consist of time, position, energy and electromagnetic field values at each step, while a typical track output group may be composed of initial and final positions and energies of tracks and information about the generation or the termination process.

%An example for a group of step outputs is listed in the following.
%
%\lstset{language=XML}
%\begin{lstlisting}
%<ks_component_group name="output_step_basics">
%	<component_member name="step_id" field="step_id" 
%					  parent="step"/>
%	<component_member name="time" field="time" 
%					  parent="step_final_particle"/>
%	<component_member name="position" field="position" 
%					  parent="step_final_particle"/>
%	<component_member name="magnetic_field" field="magnetic_field" 
%					  parent="step_final_particle"/>
%	<component_member name="electric_potential" field="electric_potential" 
%					  parent="step_final_particle"/>
%	<component_member name="kinetic_energy" field="kinetic_energy_ev" 
%					  parent="step_final_particle"/>
%	<component_member name="polar_angle_to_b" field="polar_angle_to_b" 
%					  parent="step_final_particle"/>
%</ks_component_group>
%\end{lstlisting}

%As already said, basically each information of the simulation to be obtained with a getter function can be used to construct a chain of output components and be %written down to disk.

\subsection{Analysis logic}
In addition to the flexibility of the output system, where the user can define exactly which variable at which interval or geometrical state should be recorded, it is also possible to apply a simple analysis logic to the output stream. This is highly useful, if for example, a user is only interested in the maximal kinetic energy of each track, but does not want to record the energy information for each step. The alternative of storing all this information and performing the analysis after the Monte Carlo is finished would increase the required disk space significantly.

The available analysis logic for output components within \textsc{Kassiopeia} allows for the determination of a minimal, maximal, or integral value. 
These output components can be used at any level, but the resulting output should be collected at least one level higher than the component it
depends. For example, the maximal value of a variable which is updated at step level should only be recorded at the track level or higher.
Additionally, there is a math output component, which can combine arbitrary output components at the same level with a user-defined function.

%In the following XML example a math output component is used to calculate the longitudinal kinetic energy component of the particle at each step, using the output components of kinetic energy and the polar angle between the particle momentum and the magnetic field. Furthermore,  a minimum output component is used to record the minimal longitudinal kinetic energy, which can be written down for each track.

%\lstset{language=XML}
%\begin{lstlisting}
%<ks_component_member name="step_final_particle"
%					 field="final_particle"
%					 parent="step"/>
%<ks_component_member name="kinetic_energy"
%					 field="kinetic_energy_ev" 
%					 parent="step_final_particle"/>
%<ks_component_member name="polar_angle_to_b" 
%					 field="polar_angle_to_b" 
%					 parent="step_final_particle"/>
%<ks_component_math name="longitudinal_kinetic_energy"
%				   term="x0*cos(x1*TMath::Pi()/180.)*cos(x1*TMath::Pi()/180.)" 
%				   component="kinetic_energy" 
%				   component="polar_angle_to_b"/>
%<ks_component_minimum name="min_long_kin_energy"
%					  component="longitudinal_kinetic_energy"/>
%\end{lstlisting}

\subsection{Writers and Readers}
Apart from a simple ASCII writer, the main writer for \textsc{Kassiopeia} is based on the binary data format of a \textsc{Root} \cite{root} file with a tree structure. For each selected group of output components of the four organization levels of run, event, track and step, a tree is created with the data objects as branches. Additional meta information is also stored which permits the correct reconstruction of the data.
For a simple analysis of the data, the files can be viewed and plotted with a few clicks using the \textsc{Root} TBrowser.
For a more advanced analysis, the user has the possibility to write their own analysis tools by linking against the \textsc{Kassiopeia} package. With the provided reader classes, saved data can be reconstructed automatically for user-friendly access.
Additionally, geometry data and simulation data such as steps and tracks can also be written using the VTK format \cite{schroeder2004visualization}, which
enables the creation of three-dimensional visualizations of geometries and tracks, as detailed in section \ref{sec:visualization}.
\section{Navigation}
\label{sec:navigation}

The task of the navigation module is to make sure that the simulation algorithm always follows the state defined by the current geometry object which houses the particle, as defined by the user in the configuration file. At the beginning of each track the navigator needs to check the location of the initial particle position and adjust the simulation algorithm accordingly. After each step the navigator checks if the current space was exited, a child space was entered, or a surface was crossed. This is done by calculating the distance to all relevant navigation geometries. As a caching system is used, these distances are not computed at each step, but only if the length of the particle's trajectory exceeds the cached limit. If a crossing of any navigation boundary is detected, the position of the final particle is adjusted to the exact geometrical position of the intersected surface.

If this surface is associated with a change of the simulation algorithm, the corresponding commands are executed and the next step of the particle is a surface step, meaning no propagation takes places and only the interaction associated with this surface is executed. Afterwards, the navigator checks if the particle has been reflected or transmitted, depending on the result of the surface interaction. Subsequently, the appropriate commands of the space associated with the particle's next step are executed.

If the crossed surface is part of a space change, without (surface) commands attached to it, the navigator will adjust the simulation algorithm according to the corresponding commands. If a space was exited, the inverse of the commands attached to this space are executed, and if a new space is entered, the commands attached to this navigation space are applied.
When entering a nested space, the particle still remains inside the encompassing outer space, therefore the inverse commands associated with
an exit condition are not executed. This ability to nest spaces requires that the user ensure the set of commands that are applied when entering a space are defined in a way which respects the spatial hierarchy (to avoid repetition or the removal of objects which are no longer active).

\begin{figure}[tbp]
 %\centering
    \includegraphics{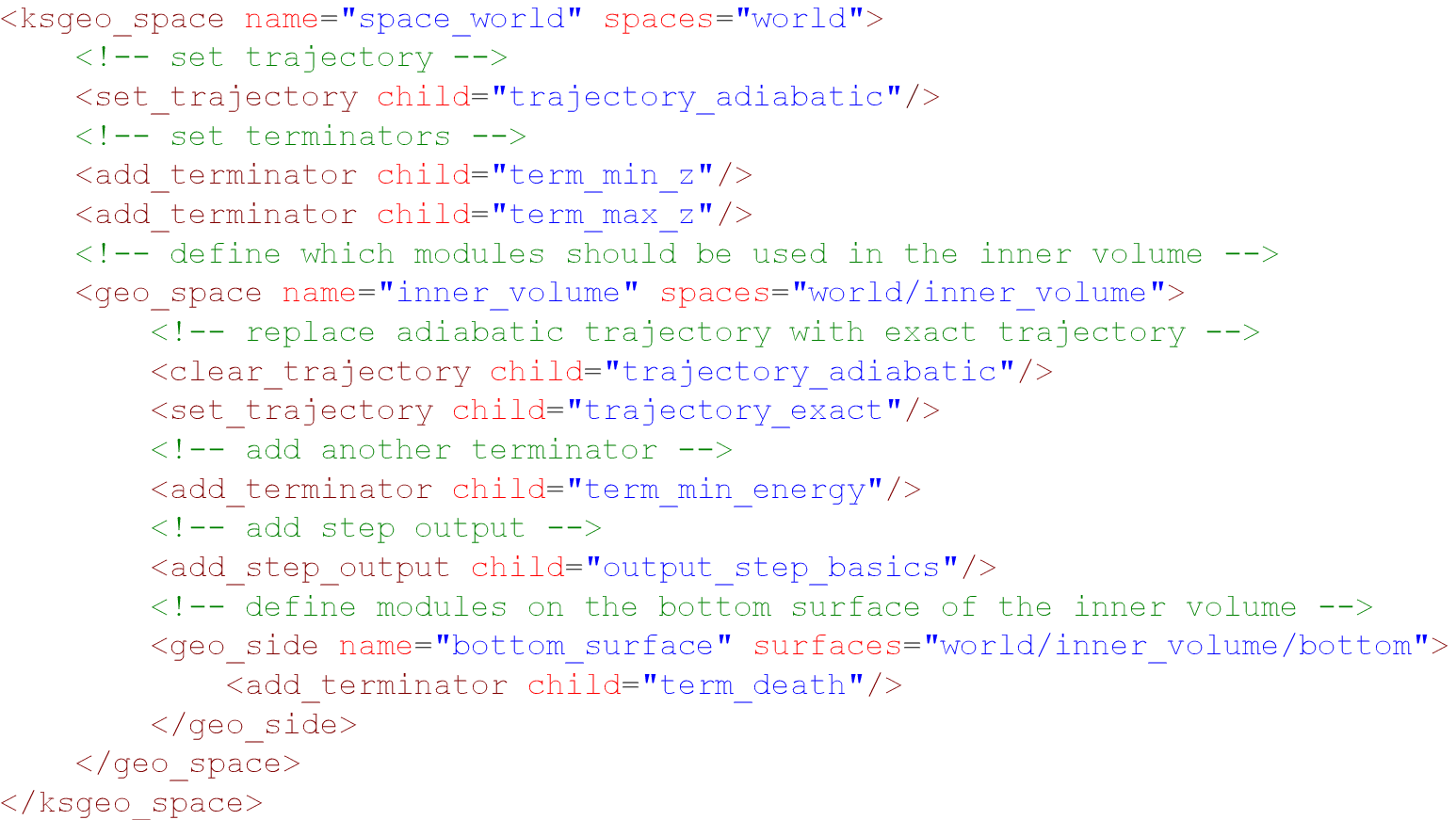}
 \caption[Example of navigation in an XML file]{Example for a navigation definition in an XML file. The shown example is explained in full detail in the text.}
\label{fig:navigation}
\end{figure}

The change of the simulation algorithm's state is achieved by commands, which can be used to modify the root classes, as detailed in section \ref{sec:design}. The definition of these commands in the XML configuration file is shown in the following example of figure \ref{fig:navigation}, where a simulation is set up in a world space with a cylindrical inner volume. When the particle enters this inner volume, the propagation calculation should switch from the adiabatic to the exact method, the step output should be activated and the particle should be terminated if it reaches the bottom surface.

%\lstset{language=XML}
%\begin{lstlisting}
%<ksgeo_space name="space_world" spaces="world">
%	<!-- add trajectory -->
%	<command parent="root_trajectory" field="set_trajectory"
%			 child="trajectory_adiabtic"/>
%	<!-- add terminators -->
%	<command parent="root_terminator" field="add_terminator" child="term_min_z"/>
%	<command parent="root_terminator" field="add_terminator" child="term_max_z"/>
%	<!-- define which modules should be used in the inner volume-->
%	<geo_space name="inner_volume" spaces="world/inner_volume">
%		<!-- replace adiabatic trajectory with exact trajectory -->
%		<command parent="root_trajectory" field="clear_trajectory" 
%				 child="trajectory_adiabatic"/>
%		<command parent="root_trajectory" field="set_trajectory" 
%				 child="trajectory_exact"/>
%		<!-- add another terminator -->
%		<command parent="root_terminator" field="add_terminator"
%				 child="term_min_energy"/>
%		<!-- add step output -->
%		<command parent="write_root" field="add_step_output" 
%				 child="output_step_basics"/>
%		<!-- define modules on the bottom surface of the inner volume -->
%		<geo_side name="bottom_surface" surfaces="world/inner_volume/bottom">
%			<command parent="root_terminator" field="add_terminator"
%					 child="term_death"/>
%		</geo_side>
%	</geo_space>
%</ksgeo_space>
%\end{lstlisting}

The example shown consists of two navigation spaces and a navigation side. The main navigation space is the world space, which possesses an attached command which activates the adiabatic trajectory within the root trajectory object, and several more commands which add terminators to the root terminator. The world space also
contains an inner volume which is of special interest to the user in the given example. In this volume, the trajectory is replaced by first defining commands for removing the old adiabatic trajectory and subsequently adding the exact trajectory to the root trajectory object. An additional terminator is added within this new space and also the step output is activated by adding an output group to the root writer. The particle should be terminated when reaching the bottom surface of the inner volume. As this particular surface is part of a navigation space it is referred to as navigation side, in this case a command is attached to it which adds a death terminator, which kills the particle track.

The available command pattern is not limited to the modification of the root classes only, but can be used to modify any object registered in the toolbox. For
example, changing the step size control of a trajectory object or adding a scattering calculator to a scattering module is easily accomplished in this paradigm.

\section{Visualization}
\label{sec:visualization}
In order to aid the user in reducing and understanding the Monte Carlo data produced by \Kassiopeia{} two visualization options are available for generating graphical output: the three-dimensional VTK visualization \cite{schroeder2004visualization} and the two-dimensional \textsc{Root} \cite{root} visualization. These two options each depend on their corresponding external libraries. Both visualization options are completely configurable in the XML configuration file, and some options include 
the ability to color track or step elements by dynamic variables, change the viewing angle, and select and plot geometric objects.
\begin{figure}[H]
	\centering
	\includegraphics[width=0.7\textwidth]{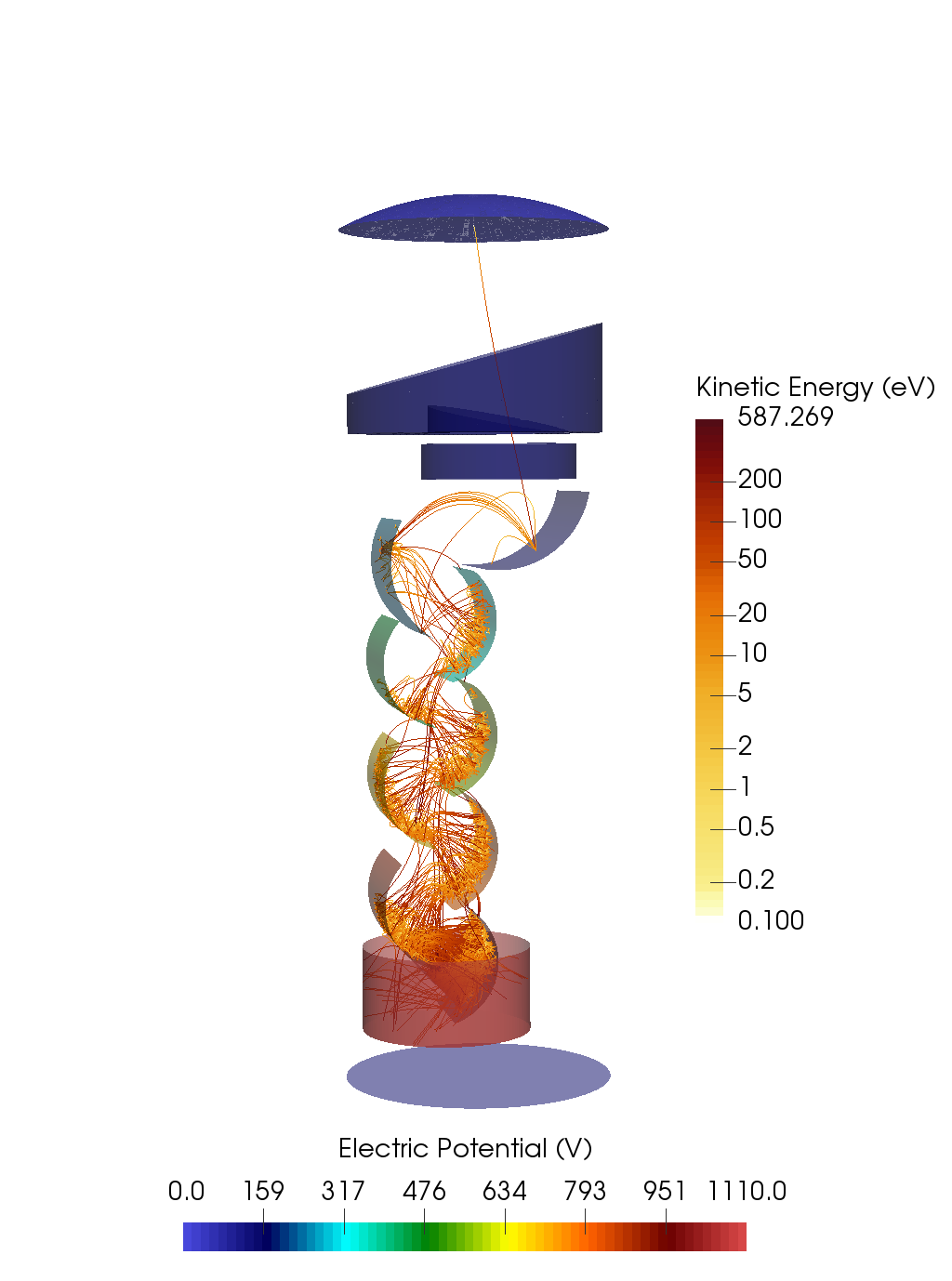}
	\caption[Example of PMT simulation]{A series of electron tracks generated from a single photo-electron event in a simplified photomultiplier tube (PMT) demonstration. The outer wall of the PMT has been removed for visualization purposes. This image was rendered using Paraview \cite{ayachit2015paraview}.}
	\label{fig:visualizaton:pmt}
\end{figure}
The first depends on the use of the external library VTK \cite{schroeder2004visualization}. As detailed in section \ref{sec:output} various data components at track and step scope can be saved to a VTK polydata file (.vtp). The user may then process the polydata files using the visualization tool of their choice. In addition, data concerning the geometry, such as the size and aspect ratio of BEM mesh elements can also be exported and visualized in this way. Figure \ref{fig:visualizaton:pmt} demonstrates some of the capabilities of the \Kassiopeia{} package and the use of the VTK output format in conjunction with the visualization software Paraview. In this example the electron tracks are colored according to their kinetic energy while the photomultiplier tube (PMT) surfaces (consisting of roughly 150K triangular elements) are colored according to their electrostatic potential.

\begin{figure}[H]
	\centering	
	\includegraphics[width=0.9\textwidth]{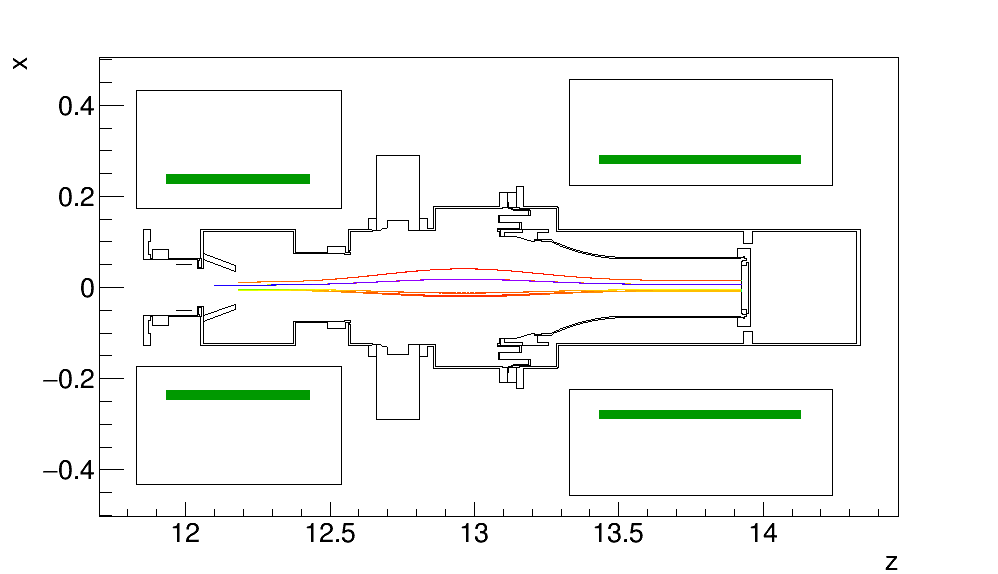}
	\caption[KATRIN Detector Simulation]{Cross section of the KATRIN Detector system with electrons simulated as entering from the left and terminating on the detector at right. The superconducting coils are shown in green providing a guiding magnetic field. The trumpet-shaped post acceleration electrode sits
	immediately before the detector and is held on +10 kV potential.}
	\label{fig:visualizaton:root}  	
\end{figure}

The second visualization option depends on the use of the external library \textsc{Root} \cite{root}, which it uses for rendering and display. This is typically used to produced a two-dimensional cross section of a selected geometry where particle tracks can be plotted and colored according to the user's preferences. In the example of figure \ref{fig:visualizaton:root}, the particle tracks are colored according to the pitch angle of the electron with respect to local magnetic field. While visually less impressive than the three-dimensional option afforded by VTK, this type of visualization can be very useful for detailed geometric investigations and is simple to use.

\section{Validation and Use}
\label{sec:validation}

Two very important issues of any scientific software are generality and correctness. Therefore it is of great importance to provide
example use cases and validate the results which \Kassiopeia{} produces. One very basic feature of the Kassiopeia developement
cycle is the existence of a set of unit and integration tests which check the basic functionality of separate and combined parts of the software. 
This tool aids in the continual development and simultaneous use of the software and is critical to keep the software in working order.

To establish the utility of \Kassiopeia{} we provide a list of published real-world use cases as follows:
 \begin{itemize}
	\item Radon induced background processes in the KATRIN pre-spectrometer: Tracking simulations are used \cite{FraenklePreSpec}.
	\item Background due to stored electrons following nuclear decays at the KATRIN experiment and its impact on the neutrino mass sensitivity:
Kassiopeia is used for stored particle tracking with interactions \cite{RadonSusanne}.
	\item Stochastic Heating by ECR as a Novel Means of Background Reduction in the KATRIN Spectrometers:
Kassiopeia is used for stored particle tracking with interactions \cite{ECR}.
	\item The KATRIN Pre-Spectrometer at reduced Filter Energy: Kassiopeia is used for tracking simulations \cite{prall}.
	\item Electromagnetic design of the large-volume air coil system of the KATRIN experiment: Magnetic Field calculation, Field line tracking \cite{aircoil:emdpaper}.
	%\item The aSPECT Experiment in Mainz uses Kassiopeia to simulate its MAC-E-Filter as well "http://www.ag-heil.physik.uni-mainz.de/186\_ENG\_HTML.php"
\end{itemize}
In addition to the aforementioned use cases, there are also several examples which provide validation 
through a direct comparison of the output of a Kassiopeia simulation
with actual experimental data. These include:
\begin{itemize}
	\item Validation of a model for radon-induced background processes in electrostatic spectrometers: Stored particle tracking with interactions \cite{RadonValidation}.
	\item Modeling of electron emission processes accompanying radon-$\alpha$-decays 
	within electrostatic spectrometers: Validation of the Radon generator \cite{RadonModel}.
	\item High voltage monitoring with a solenoid retarding spectrometer of the KATRIN experiment: Magnetic field calculation, Field line tracking \cite{moni:erhard}.
	\item Investigation of the passage of electrons from vacuum into the active volume of a p-i-n diode charged particle detector: Electron tracking and diffusion in silicon. \cite{wall2014dead}.
	\item Project8, an experiment trying to measure the neutrino mass by taking tritium$\beta$ spectra through detection of cyclotron radiation relies on Kassiopeia to model the radiative losses by electron's in a magnetic trap.
\end{itemize}

\begin{figure}
	\centering	
	\includegraphics[width=0.9\textwidth]{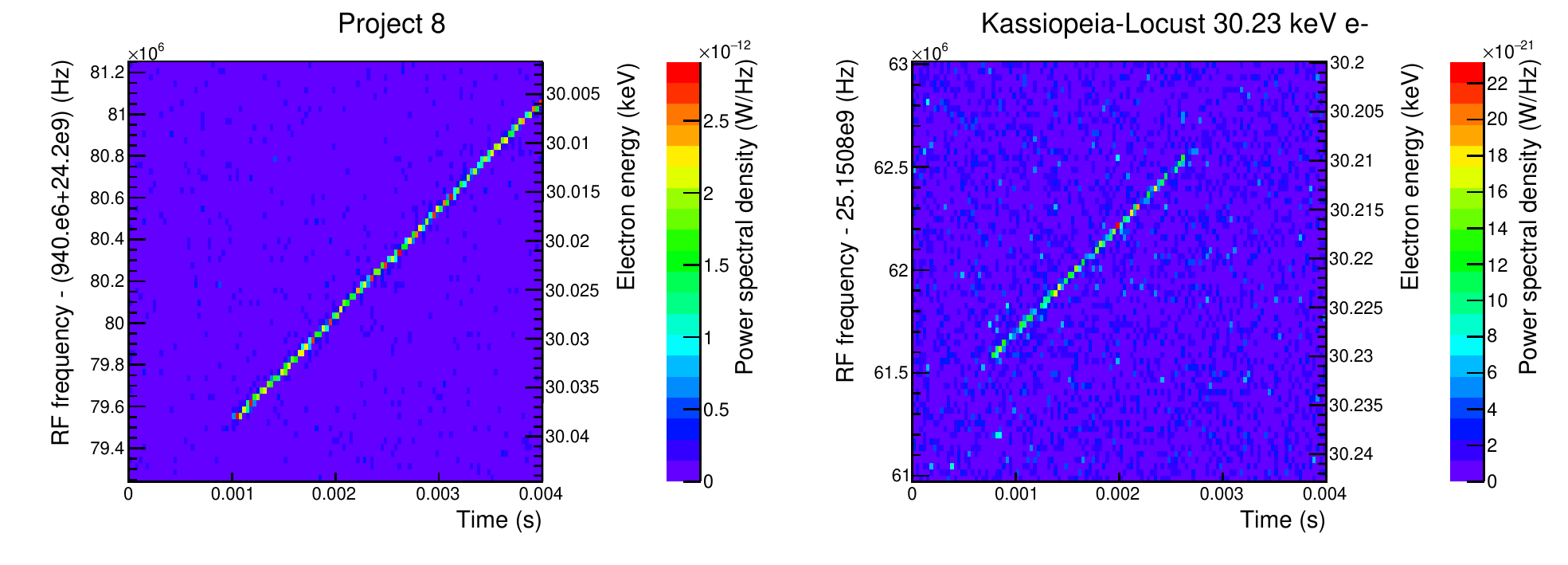}
	\caption[Comparison of radiative losses from trapped electrons.]{Side by side comparison showing cyclotron radiation from a single electron measured in the Project 8 experiment (left) and that from a 30.23 keV electron simulated in Kassiopeia (right).  On the right, the power in the simulated noise background is 10$\times$10$^{-22}$ W/Hz.  In the left panel the measured noise background is expected to be similar or lower than the above due to recent hardware improvements.  The plots demonstrate that the cyclotron radiation energy losses calculated by Kassiopeia are consistent with those observed in an empirical measurement.}
	\label{fig:validation:project8}
\end{figure}

One particularly interesting comparison between \Kassiopeia{} and experimental data is provided by the Project 8 experiment \cite{asner2015single}, the results
of which are shown in figure \ref{fig:validation:project8}. In this example a magnetic field of approximately 1.0 T is modelled in Kassiopeia.  A trapping region is implemented by distorting the main field by -40 G in a parabolic shape which is similar to that used by the Project 8 experiment.  Electrons with 
pitch angles, $\phi$, where $|\phi - 90^{\circ}| < 6^{\circ}$, are trapped magnetically for up to 1 ms.  
Their tracks are calculated in Kassiopeia using adiabatic trajectory integrated
by an 8-th order Runge-Kutta method with a stepsize limited to 1.125 cyclotron periods.  The electrons move back and forth in the trap 
at a frequency near 100 MHz while circling the field lines in cyclotron motion. Further demonstrating the flexibility 
of \Kassiopeia{} and the ease by which it may be extended, the simulated energy losses calculated by \Kassiopeia{} are sampled at 2 GHz 
and passed to an external software package called Locust which simulates the voltages induced in an antenna and receiver.

The right panel of Figure \ref{fig:validation:project8} shows the power from a simulated 30.23 keV electron located at the magnetic trap minimum with pitch angle 90$^{\circ}$. The slope of the track is in agreement with the corresponding empirical measurement from the Project 8 experiment in the left panel, confirming the calculations of energy loss.  The starting frequencies of the tracks agree to within 200 eV which is consistent with absolute uncertainty on the field at the magnetic minimum of the trap.

Complementary to the comparison to real world data, some simple physical test cases which have analytical reference solutions have been incorporated into
several stand-alone binaries. For example, the charge density calculation can be performed on several simple geometries such as
a spherical or unit-cube capacitor, which shown the field solvers are both accurate and scalable. For additional results discussing the validity and accuracy of 
the field computation the reader is referred to \cite{phd:corona} and \cite{barrett:2016}.  

\begin{figure}
	\centering	
	\includegraphics[width=0.9\textwidth]{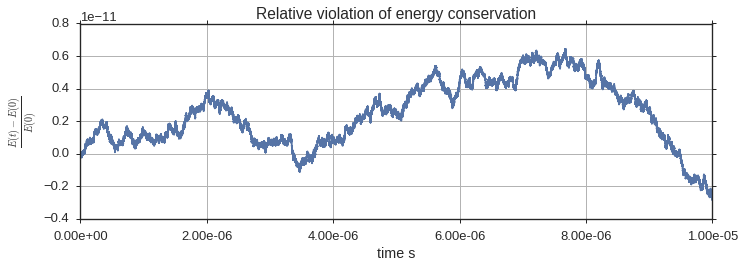}
	\caption[violation of energy conservation]{The relative violation of energy conservation for an electron stored over $\approx10~\mu \mathrm{s}$. This
		corresponds to roughly 1.7 million cyclotron periods in a perfect static quadrupole trap with a magnetic field of 1~T and a Voltage of 10~V.
		This calculation was performed using the exact tracking method with a stepsize limited to less than $10^{-11}$s. A full description is shipped with \Kassiopeia{} in the corresponding configuration file QuadrupoleTrapSimulation.xml, which enables any user to re-run this reference simulation.}
	\label{fig:validation:quadrupole}
\end{figure}

In addition, to test a complete case involving particle tracking, the simulation of a (static) quadrupole Penning trap
has been done. This test case was chosen for several reasons. It exhibits long storage times of charged particles which is a prominent use case for \Kassiopeia{} and allows one to evaluate its performance in this task. Its fields and also the trajectories of stored particles are well described analytically which makes it possible to evaluate tracking and field computation independently against known solutions. On the other hand this system is not trivial, in real world applications it can be affected by electric and magnetic field disturbances which may be further explored in simulation. For this particular example, a plot showing the degree of energy conservation violation is shown in figure\ref{fig:validation:quadrupole}.

Similar to the law of energy conservation another invariance theorem is very suitable to assess the numerical errors in this particular system. The Brown-Gabrielse theorem\cite{BrownGabrielse82} states
\begin{eqnarray}
\omega_m^2+\omega_z^2+\omega_{c'}^2 = \omega_c^2.
\end{eqnarray}  
It connects the frequencies observed in a quadrupole trap to the free cyclotron frequency $\omega_c = \frac{e}{m_e\gamma}B$ where $\omega_m$ is the magnetron frequency, $\omega_z$ is the axial frequency and $\omega_{c'}$ is the cyclotron frequency each as observed from the motion of the particle in the trap. For the same simulation used in figure \ref{fig:validation:quadrupole} a relative error of 
\begin{eqnarray}
\sqrt{\frac{\omega_m^2+\omega_z^2+\omega_{c'}^2}{\omega_c^2}}-1 = -1.86\cdot 10^{-8}
\end{eqnarray}
is observed. Hereby the frequencies were determined by determining the length of half-periods for the y- and z-components of the trajectory. The gamma factor entering the calculation of $\omega_c$ was derived from the mean kinetic energy of the electron of about $0.8~$eV whereas neglecting this leads to an error of $1.5\cdot 10^{-6}$. A commented Python notebook containing the analysis to reproduce these results is shipped with \Kassiopeia.
\begin{figure}
	\centering	
	\includegraphics[width=0.9\textwidth]{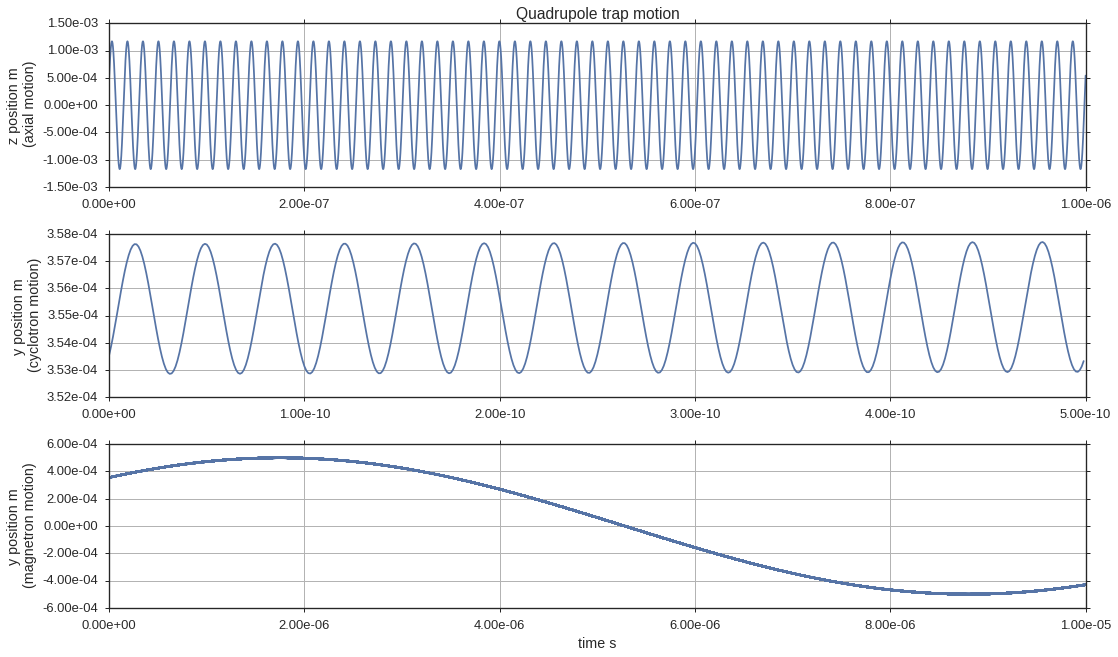}
	\caption{From the motion of the electron in the trap the frequencies are derived by calculating the mean time between two zero crossings. For the magnetron motion  $y=a_m\cdot\cos\left(\omega_m\cdot t+\phi_m\right)$ was fitted to the data and afterwards subtracted to allow for proper detection of zero-crossings for the cyclotron motion. The top plot displays the motion in z direction, below that the cyclotron motion in the y component and at the bottom the magnetron motion are shown. The data is generated by the same QuadrupoleTrapSimulation.xml as in figure\ref{fig:validation:quadrupole}} 
\label{fig:validation:gabrielse}
\end{figure}

\section{Conclusion}
\label{sec:conclusion}

We have presented the \Kassiopeia{} particle tracking framework developed within the \Katrin{} collaboration. It was design to enable the fast and accurate computation of three-dimensional and axial-symmetric static electromagnetic fields created by complex electrode and magnet geometries using \KEMField{}'s fast multipole and zonal harmonic methods respectively. In these fields particles can be tracked using various Runge Kutta integrators (available up to 8th order) over extended periods of time using either an adiabatic approximation or by exactly solving the full Lorentz equation. The effects of synchrotron radiation can be taken into account in both cases. Furthermore, interaction models for electrons scattering on gaseous molecular hydrogen as well as helium, argon and in silicon detectors are available. A major feature of the software is its flexibility that stems from the XML configuration language that is used to describe all aspects of a simulation. If the pre-existing modules are sufficient for an end-user no C++ code needs to be written and even novices can quickly devise
complex and interesting simulations. The incorporation of new effects and models does require C++ knowledge but their inclusion
is made comparatively simple given the modular structure of the software.
By now it is heavily used within the \Katrin{} collaboration to study all aspects of the corresponding experiment. Beyond \Katrin{} it is used by the aSPECT \cite{simson2009measuring} and Project8 \cite{asner2015single} collaborations. Therefore, we believe that this tool will be useful to a wider scientific community and hope for new users. The source code along with instructions for installing dependencies, compilation, and the first steps of configuration is available for download from: \url{https://github.com/KATRIN-Experiment/Kassiopeia}

The continued development of \Kassiopeia{} will be on-going and it is expected that many improvements and additional features will be available in
the future. Upcoming features will include the ability to track neutral particles with spin, and particles which have an internal quantum state.
Futhermore, the field code will eventually incorporate additional integration techniques \cite{gluck2016electric} to improve numerical stability and may be extended to non-static problems using time or frequency domain BEM techniques. 

This material is based upon work supported by the U.S. Department of Energy, Office of Science, Office of Nuclear Physics under Award Numbers
FG02-97ER41041 and DE-FG02-06ER-41420. In addition, this work was supported by the German BMBF (05A14VK2), HAP, KHYS, and KCETA.

\clearpage
\bibliographystyle{iopart-num}
\bibliography{kassiopeia}

\end{document}